\documentclass[superscriptaddress,aps,prd,twocolumn,showpacs,nofootinbib]{revtex4-1}
\usepackage{etex}
\usepackage{amsmath,amssymb,amsthm}
\usepackage{ragged2e}
\usepackage[colorlinks=true,citecolor=blue,urlcolor=blue]{hyperref}
\usepackage{verbatim}   
\usepackage{color}      
\usepackage{subfigure}  
\usepackage{hyperref}   
\usepackage{mathrsfs}
\usepackage{booktabs}
\usepackage[font=footnotesize,labelfont=bf]{caption} 
\usepackage{wrapfig}
\usepackage{bm}
\input{epsf}
\usepackage{float}
\usepackage{graphicx}
\usepackage{dcolumn}
\usepackage{bm}
\usepackage{hyperref}
\hypersetup{colorlinks,citecolor=blue}
\usepackage{color}
\usepackage{epstopdf}
\begin{document}

\preprint{APS/123-QED}

\title{ Neutron Stars with realistic EoS in $f(R)$ theories of  gravity }

\author{K Nobleson}
\affiliation{%
 Department of Physics, BITS Pilani, Hyderabad Campus,\\ Jawaharnagar, Hyderabad 500078, India
}%


\author{Amna Ali}
\affiliation{
 Department of Mathematics, Jadavpur University,\\ Kolkata 700032, West Bengal, India
}%
\author{Sarmistha Banik}
\affiliation{%
Department of Physics, BITS Pilani, Hyderabad Campus,\\ Jawaharnagar, Hyderabad 500078, India
}%


\date{\today}

\begin{abstract}
In this paper, we examine neutron star structure in perturbative $f(R)$ gravity models with realistic equation of state. We obtain mass-radius relations in two gravity models of the form $f_{1}(R)=R+ \alpha R(e^{-R/R_0}-1)$ and $f_{2}(R)=R+\alpha R^2$. For this purpose, we consider NS with several nucleonic as well as strange EoSs generated in the framework of relativistic mean field models. The strange particles in the core of NS are in the form of $\Lambda$ hyperons and quarks, in addition to the nucleons and leptons. The M-R relation of the chosen EoSs lies well within the observational limit in the case of GR. We show that these EoSs provide the most stringent constraint on the perturbative parameter $\alpha$ and therefore can be considered as important experimental probe for modified gravity at astrophysical level.

\end{abstract}

\maketitle


\section{\label{sec:intro} Introduction}
The Einstein's theory of General Relativity(GR)\cite{GR} has been the most successful theory of the twentieth century which gives the most plausible description of the Universe at the large scales. Along with the Standard Model of particle physics, it forms the "Grand Unified Theory" from which all physics may be derived. The greatest insight of GR is that it describes gravity, one of the fundamental forces, not as a force but as an inherent property of space-time which arises by considering the Universe as a geometric manifold. Still, ever since the inception of the theory, people looked for alternative theories of gravity. The main reason, then, was the lack of experimental confirmation of its strong-field predictions until 60 years after its formulation. Also, quantitative hypothesis of the theory was against the predictions from Newtonian cosmology, conceptual difficulties in quantizing Einstein's theory, and astrophysical observations suggested that GR may require modifications. 

In recent years, a great interest has been made to study the modified gravity theories, mainly with the goal of finding physical explanations to the accelerated expansion of the universe, which has been confirmed by several independent observations \cite{perlmutter1999measurements}. The simplest solution consistent with the observations of recent cosmic acceleration is the cosmological constant $\Lambda$. However, the field theoretic understanding of $\Lambda$ is far from being satisfactory and its magnitude is significantly small from what is expected, which leads to the well-known {\it coincidence} and {\it fine-tuning} problems. To address the said problem, two most popular approaches have been adopted $-$ either introduction of an additional scalar field \cite{tsujikawa2006dynamics}, or the related approach of replacing the Ricci Scalar $R$ in Einstein-Hilbert action with a general function of the Ricci scalar $f(R)$ \cite{carroll2004cosmic}. Within both the frameworks, the additional scalar degree of freedom can be tuned to mimic the cosmological constant or any type of viable cosmological evolution at any cosmological scales. 

The modified theories of gravity are well motivated and consistent with the observational data without requiring any addition of ad hoc component of matter, which eluded detection so far. However, such modifications should undergo all the astrophysical tests to be accepted or refuted. Majority of these models are verified for their feasibility by undergoing the solar system test and laboratory tests \cite{ozel2016masses}. Here the gravitational field is substantially weak and has been subject to numerous experimental tests which confirms the accuracy of GR on the weak gravitational background. But any consistent theory of gravity, classical or modified, should also be equally applicable to the strong gravity regime. The compact stars, such as neutron stars (NS), provide a good platform to study the behaviour of strong gravity. 

In $f(R)$ theories of gravity, a set of modified Tolman-Oppenheimer-Volkoff (TOV) equations describe a spherically symmetric mass distribution under hydrostatic equilibrium. It has been shown in \cite{babichev2010relativistic}, that in scalar tensor theories of gravity these modified TOV solutions reproduce the correct behaviour at the weak gravity limit. To solve the TOV equations or the modified TOV equations, we need an equation of state (EoS) characterizing the relation between the internal pressure and the energy density in a NS \cite{singh2020compact}. A polytropic EoS is of the form of $P=K\epsilon^\Gamma$ where K is a constant and $\Gamma$ is the adiabatic index. Equilibrium solution of an approximate polytropic EoS with $\Gamma \simeq 2$ yields a somewhat good agreement with the NS mass-radius observations.  However, one needs a better EoS to provide the physical description of the NS matter. The  mass-radius relations for NS in modified gravity using a polytropic EoS has been derived by following the perturbative approach \cite{cooney2010neutron}. Such an approach is well motivated and is formulated to treat the complexity of field equations by considering the correction terms to GR as only the next to the leading order terms in a larger expansion. The same approach was used for a set of realistic EoS for NS in various f(R) gravity models \cite{Arapo_lu_2011, deliduman2012neutron, orellana2013structure, Astashenok, Capozziello, Sbisa}. In this paper, we use the same approach to further examine the existence and properties of NS with realistic EoS in the context of f(R) models, namely $f_{1}(R)=R+ \alpha R(e^{-R/R_0}-1)$ and $f_{2}(R)=R+\alpha R^2$ and obtain the perturbative constraints. 

The realistic EoSs of the dense matter relevant to NS are constrained by the empirical values of the nuclear matter properties at normal nuclear matter density ($\rho_0\sim 2.4 \times 10^{17}kg/m^3$); specifically, the binding energy, compressibility, effective mass of nucleons, symmetry energy, and its slope. However, the maximum mass configurations can often reach the supra-nuclear density ($>\rho_0$). But the high-density nature of the nuclear matter is still uncertain owing to our restricted knowledge above the saturation density. Several models exist to construct the EoS at this density, which are then adjusted to nuclear experimental results and astrophysical observations. 

Additional degrees of freedom other than neutrons, protons, and electrons are believed to populate the high-density core of a compact star. The presence of exotic particles softens the EoS, which in turn yields a lower maximum mass NS. The observational mass is an useful tool to constrain the EoS. The first major mass measurement of two massive pulsars of masses $1.928\pm 0.017$ \cite{fonseca2016nanograv} and $2.01\pm0.04 M_{\odot}$ \cite{antoniadis2013t} ruled out the soft EoSs comprising of additional strange particles, which fail to support such a massive star. A few exotic EoSs with hyperons, antikaons, and quarks, however, remain compatible with the above observations. The first detection of gravitational waves from a binary NS merger, GW170817, has set some upper bound on the tidal deformibility ($\Lambda$) \cite{abbott2017gw170817, abbott2018gw170817}, which on the other hand has been used by many authors to constrain the NS radius. For example, a limit of $9.8 < R < 13.2$ km is given in Ref \cite{raithel2019constraints}, which is consistent with the radius measurement from X-ray observations\cite{ozel2016masses1}. NASA's Neutron star Interior Composition Explorer (NICER) has recently published mass and size data for the solitary pulsar J0030+0451, located 1,100 light-years away in the constellation Pisces\cite{2019ApJ...887L..22R}. Researcher have estimated its mass and radius around $1.3-1.4M_{\odot}$ and radius $\sim 12.7-13 $ kilometers \cite{2019ApJ...887L..22R, Miller_2019}. In this paper, we use the relativistic EoSs which are compatible with this mass-radius observations.
 
The paper is organised as follows. In section II, we review the field equations of $f(R)$ modified gravity model considering its perturbative form. Assuming perturbative forms of metric functions and the hydrodynamical quantities, we obtain the modified TOV equations. In section III, we present an overview of the EoSs used in this work. In section IV, the modified TOV equations are solved numerically for various forms of the EoSs, the functional forms of $f(R)$, and various values of the perturbation parameter $\alpha$. Finally, in the discussion section, we comment on the results of numerical study and on the significance of the scale of the perturbation parameter $\alpha$ for the two  modified gravity models, namely $f_1(R)=R+ \alpha R(e^{-R/R_0}-1)$ and $f_2(R)=R+\alpha R^2$. 

\section{\label{sec:mTOV} $f(R)$ Gravity and the Modified TOV Equations}
The $4$-dimensional action of $f(R)$ gravity is the simplest generalization of the Lagrangian in the Einstein-Hilbert action as:
\begin{eqnarray}\label{action}
S=\frac{1}{16\pi}\int d^4x \sqrt{-g}f(R) + S_{{\rm m}}
\end{eqnarray}
where $g$ denotes the determinant of the metric $g_{\mu\nu}$, $R$ is the Ricci scalar, and $ S_m$ is a matter action. We have set $G$ and $c$ to $1$ in the action. Following the metric formalism of the action and adopting a perturbative approach as in \cite{cooney2010neutron}, we choose the function $f(R)$ such that the deviation from GR is small and is parameterized by $\alpha$ as:
\begin{eqnarray}\label{fR}
    f(R)=R+\alpha h(R)+\mathcal{O}(\alpha ^{2})
\end{eqnarray}
where $h(R)$ is an arbitrary function of $R$ and $\mathcal{O}(\alpha ^{2})$ denotes the possible higher-order corrections in $\alpha$. Considering the above form of $f(R)$, the field equations are derived by the variation of action Eq. \ref{action} with respect to the metric tensor $g_{\mu\nu}$ 
\begin{eqnarray}\label{field}
8\pi T_{\mu \nu } &=& (1+\alpha h_{R})G_{\mu\nu} -\frac{1}{2}\alpha(h-h_{R}R)g_{\mu \nu}\nonumber
\\
& &-\alpha (\nabla _{\mu }\nabla _{\nu }-g_{\mu \nu }\Box )h_{R}
\end{eqnarray}
where $h_R=\frac{dh}{dR}$ and $T_{\mu \nu }=-\frac{2}{\sqrt{-g}}\frac{\delta S_m}{\delta g^{\mu \nu}}$ is the energy-momentum tensor of the matter fields. The trace of the above equation is 
\begin{eqnarray}\label{trace}
	8\pi T = R - \alpha\left[h_R R - 2h+ 3\Box h_R\right] +O(\alpha^2).
\end{eqnarray}
As we are interested in spherically symmetric solutions of these field equations inside a NS, we choose a spherically symmetric metric of the form:
\begin{equation}\label{metric}
    ds^2= -e^{2\phi}dt^2 +e^{2\lambda}dr^2 +r^2 (d\theta^2
    +\sin^2\theta d\phi^2).
\end{equation}
Considering the energy-momentum tensor of the perfect fluid,  ``tt'' and ``rr'' components of field, Eq. \ref{field} becomes
\begin{eqnarray}
-8\pi \rho &=& -r^{-2} +e^{-2\lambda}(1-2r\lambda')r^{-2} \nonumber \\
    && +\alpha h_R[-r^{-2} +e^{-2\lambda}(1-2r\lambda')r^{-2}]   -\frac12\alpha(h-h_{R}R) \nonumber \\ &&+e^{-2\lambda}\alpha[h_R'r^{-1}(2-r\lambda')+h_R''] \label{f-tt}
    \end{eqnarray}
    \begin{eqnarray}
  8\pi P &=& -r^{-2} +e^{-2\lambda}(1+2r\phi')r^{-2} \nonumber \\
    &&+\alpha h_R[-r^{-2} +e^{-2\lambda}(1-2r\lambda')r^{-2}] \nonumber \\
             && -\frac12\alpha(h-h_{R}R) +e^{-2\lambda}\alpha h_R'r^{-1}(2+r\phi') \label{f-rr}
\end{eqnarray}
where $'=d/dr$, i.e., it denotes derivative with respect to radial distance $r$. We assume a Schwarzchild solution for the exterior region, therefore it is convenient to define the change of variable 
\begin{equation}
\label{mass}
    e^{-2\lambda}=1-\frac{2 M}{r}
    \end{equation}
where M denotes gravitational star mass which relates to the density $\rho$ as:
   $ M=8\pi \int \rho(r) r^2 dr.$
Taking the derivative of $M$ with respect to $r$, one obtains:
\begin{eqnarray}\label{dMa/dr}
    \frac{dM}{dr}=\frac{1}{2}[ 1-e^{-2\lambda}(1-2r\lambda')]
\end{eqnarray}
We define dimensionless variables,
$$
M\rightarrow m M_{\odot},\quad r\rightarrow r_{g}r, \quad \rho\rightarrow\rho M_{\odot}/r_{g}^{3},$$ 
$$ \quad P\rightarrow p M_{\odot}c^{2}/r_{g}^{3}, \quad R\rightarrow {R}/r_{g}^{2},$$ where $M_{\odot}$ is the Sun mass and $r_{g} = GM_{\odot}/c^{2} = 1.47473$ km.  Substituting  Eq.\ref{dMa/dr} in Eq.\ref{f-tt} we get the first TOV equation as:
\begin{widetext}
\begin{equation}\label{TOV-1}
\left(1+\alpha r_{g}^{2} h_{{R}}+\frac{1}{2}\alpha r_{g}^{2} h'_{{R}} r\right)\frac{dm}{dr}=4\pi{\rho}r^{2}-\frac{1}{4}\alpha r^2 r_{g}^{2}\left[h-h_{{R}}{R}-2\left(1-\frac{2m}{r}\right)\left(\frac{2h'_{{R}}}{r}+h''_{{R}}\right)\right]
\end{equation}
\begin{eqnarray}\label{TOV-2}
8\pi p\!=\!\!-2\left(1+\alpha r_{g}^{2}h_{{R}}\right)\frac{m}{r^{3}}-\!\!\left(1-\!\frac{2m}{r}\right)\!\!\left[\frac{2}{r}(1+\alpha r_{g}^{2} h_{{R}})+\alpha r_{g}^{2} h'_{{R}}\right]\frac{1}{({\rho}+p)}\frac{dp}{dr}-
\!\frac{1}{2}\alpha r_{g}^{2}\!\left[h-h_{{R}}{R}-4\!\left(1-\frac{2m}{r}\right)\!\frac{h'_{{R}}}{r}\right]
\end{eqnarray}
\end{widetext}
The second TOV equation above can be obtained by substitution of the derivative $d\phi/dr$ in Eq.\ref{f-rr} from the hydrostatic equilibrium equation
\begin{equation}\label{hydro}
\frac{dP}{dr}=-(\rho
    +P)\frac{d\phi}{dr}
   \end{equation}
 The conservation equation of energy-momentum of a perfect fluid, $\nabla^\mu T_{\mu\nu}=0$, yields  Eq.\ref{hydro}. 

 For $\alpha=0$, Eqs. \ref{TOV-1} and \ref{TOV-2}  reduce to
\begin{eqnarray}
\frac{dm}{dr}&=&4\pi\rho r^{2}\\
\frac{dp}{dr}&=&-\frac{4\pi p r^{3}+m}{r(r-2m)}\left(\rho+p\right)
\end{eqnarray}
Note that these are the original dimensionless TOV equations in case of GR. These equations can be solved numerically for a given EoS $p=f({\rho})$ and initial conditions $m(0)=0$ and ${\rho}(0)={\rho}_{c}$, the central density.

For non-zero $\alpha$, one needs the third equation for the Ricci curvature scalar. R can be deduced by the trace of field equation given by Eq. \ref{trace}. In dimensionless variables, it can be written as:
\begin{widetext}
\begin{eqnarray}\label{TOV-3}
8\pi({\rho}-3p)\!=\!R-\!3\alpha r_{g}^{2}\left[\Big\{\frac{2}{r}-\!\frac{3m}{r^{2}}-\!\frac{dm}{rdr}-\!\!\left(1-\frac{2m}{r}\right)\frac{dp}{(\rho+p)dr}\Big\}\frac{d}{dr} +
\!\!\left(1-\frac{2m}{r}\right)\frac{d^{2}}{dr^{2}}\right] h_{R}  
-\alpha r_{g}^{2} h_{{R}}{R}+2\alpha r_{g}^{2} h
\end{eqnarray}
\end{widetext}
It is to be noted that the term $\alpha r_{g}^{2} h(R)$ is a dimensionless function. We need to add the EoS for matter inside star to solve the Eqs. \ref{TOV-1}, \ref{TOV-2}, and \ref{TOV-3}. The solution of above TOV equations can be achieved by using the perturbative approach, where the density, pressure, mass, and curvature can be expanded as
\begin{eqnarray}
p=p^{(0)}+\alpha p^{(1)}+...,\quad \rho=\rho^{(0)}+\alpha \rho^{(1)}+...,\nonumber \\
m=m^{(0)}+\alpha m^{(1)}+...,\quad R=R^{(0)}+\alpha R^{(1)}+....
\end{eqnarray}
Here $\rho^{(0)}$, $p^{(0)}$, $m^{(0)}$, and $R^{(0)}$  satisfy the standard TOV equations in GR assumed at zeroth order. Terms containing $h_{R}$  are already first order in $\alpha$, therefore all such terms should be evaluated at  ${\mathcal O}(\alpha)$ order. With $m=m^{(0)}+\alpha m^{(1)}$ and $p=p^{(0)}+\alpha p^{(1)}$, the modified TOV equations are as follows:
\begin{widetext}
\begin{eqnarray}\label{TOV1}
\frac{dm}{dr}&=&4\pi\rho r^2-\alpha r^{2}\left[4\pi \rho^{(0)}h_{R}+\frac{1}{4}\left(h-h_{R}R\right)\right]
+\frac{1}{2}\alpha\left[\left(2r-3m^{(0)}-4\pi\rho^{(0)}r^{3}\right)\frac{d}{dr}+r(r-2m^{(0)})\frac{d^{2}h_R}{dr^{2}}\right]
\end{eqnarray}

\begin{equation}\label{TOV2}
\frac{r-2m}{\rho+p}\frac{dp}{dr}=4\pi r^2 p+\frac{m}{r}-\alpha r^2\left[4\pi p^{(0)}h_{R}+\frac{1}{4}\left(h-h_{R}R\right)\right] - \alpha \Big(r-3m^{(0)}+2\pi p^{(0)}r^{3}\Big)\frac{dh_{R}}{dr}
\end{equation}
\end{widetext}

The Ricci curvature scalar, in terms containing $h_{R}$ and $h$, has to be evaluated at  ${\mathcal O}(1)$ order, i.e.
\begin{equation}
R \thickapprox R^{(0)}=8\pi(\rho^{(0)}-3p^{(0)})
\end{equation}
In this framework of  perturbative approach, we do not consider the curvature scalar as an additional degree of freedom since its value is fixed by the above relation. As in general relativity, the modified TOV Eqs. \ref{TOV1} and \ref{TOV2}, can be solved numerically for various functional forms of $h(R)$. We note that perturbation expansion parameter $\alpha$ introduces a new scale into the theory. Further, by choosing a realistic EoS, we compute mass-radius relation for various values of $\alpha$, thereby placing a bound on $\alpha$ for perturbative $f(R)$ gravity models viz: $f_1(R)=R+ \alpha R(e^{-R/R_0}-1)$ and $f_2(R)=R+\alpha R^2$.  

\section{Neutron Star Model and Astrophysical Constraints}
The interior temperature of the NS is low  compared to the Fermi energy of the constituent fermions. Hence the dense matter relevant to the NS core can be modeled in terms of zero temperature EoS. We consider a set of relativistic EoSs that are permissible by the tidal deformability constraint $\Lambda_{1.4} \simeq 800$, inferred from the first analysis of GW170817 event (for NS of mass $M= 1.4M_{\odot}$) \cite{abbott2017gw170817}. They also obey the mass-radius constraints from astrophysical observations \cite{fonseca2016nanograv, antoniadis2013t, demorest5788two}. We adopt the Walecka model \cite{WALECKA1974491} to construct the EoS. Originally, the interaction among the baryons in the Walecka model was described via the exchange of scalar-isoscalar $\sigma$ meson and vector-isoscalar $\omega$-meson. Later, vector-isovector $\rho$ mesons were added to take care of isospin symmetry. Non-linear scalar self-interactions and $\delta$ mesons were introduced for the high density behaviour. The equations of motion are solved where the meson fields are replaced by their mean values. Hence they are called as Relativistic mean field (RMF) models. The NS matter is charge neutral and is in $\beta$-equilibrium. Also, total baryon number is conserved.  In these models, the coupling constants are fitted to the binding energies and charge radii of finite nuclei and/or to the nuclear matter properties at the saturation density. There is another prescription, where the high-density behaviour is considered through the density dependence of the meson-baryon couplings.  We broadly divide the EoS into two categories: A) nucleons only B) nucleons with additional strange components.
\begin{figure*}[!t]
\begin{minipage}[b]{0.45\textwidth}
  \includegraphics[width=\linewidth, height=10cm]{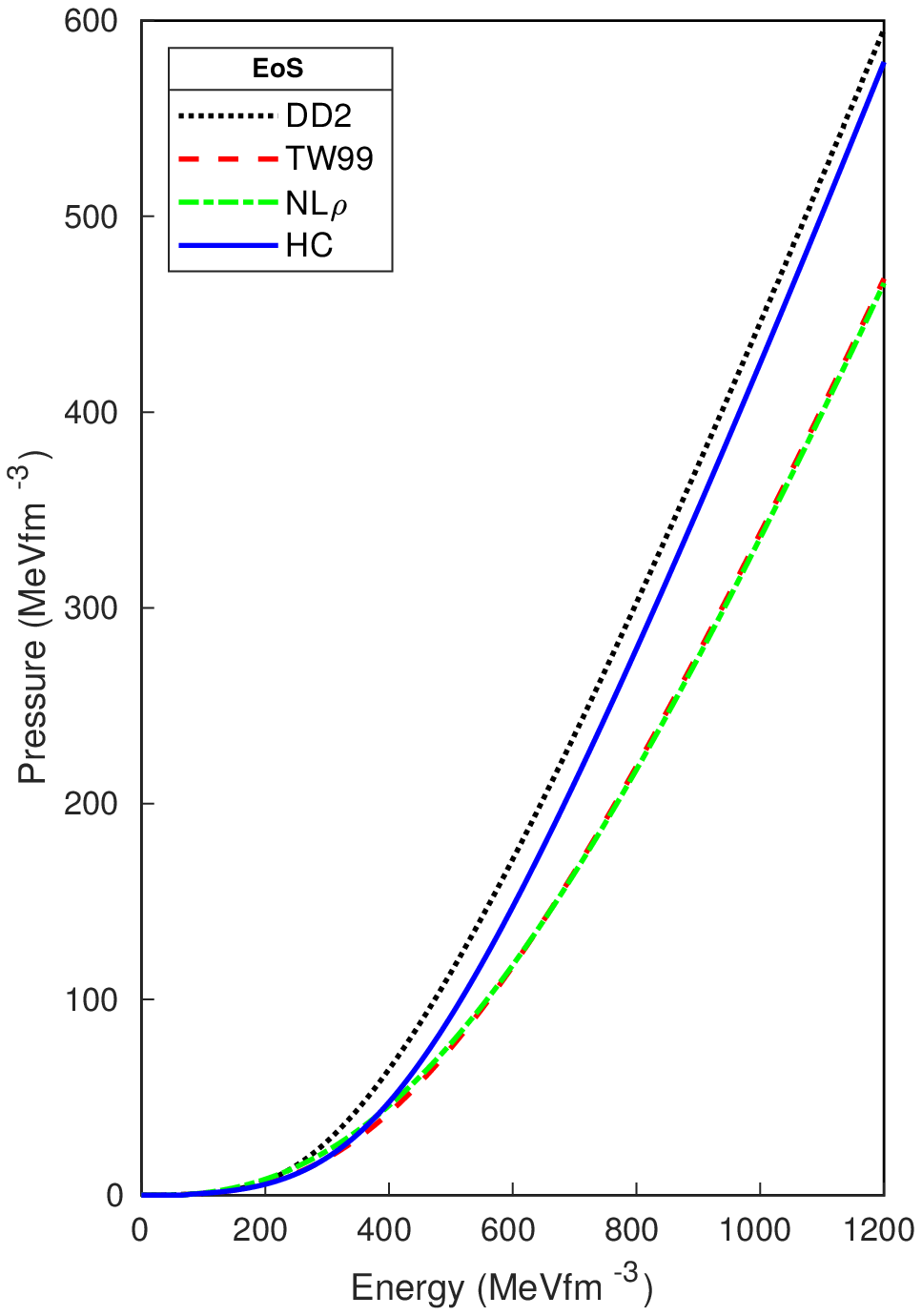}
  \mbox{(a)}
  \end{minipage}\hfill
\begin{minipage}[b]{.45\textwidth}
  \includegraphics[width=\linewidth, height=10cm]{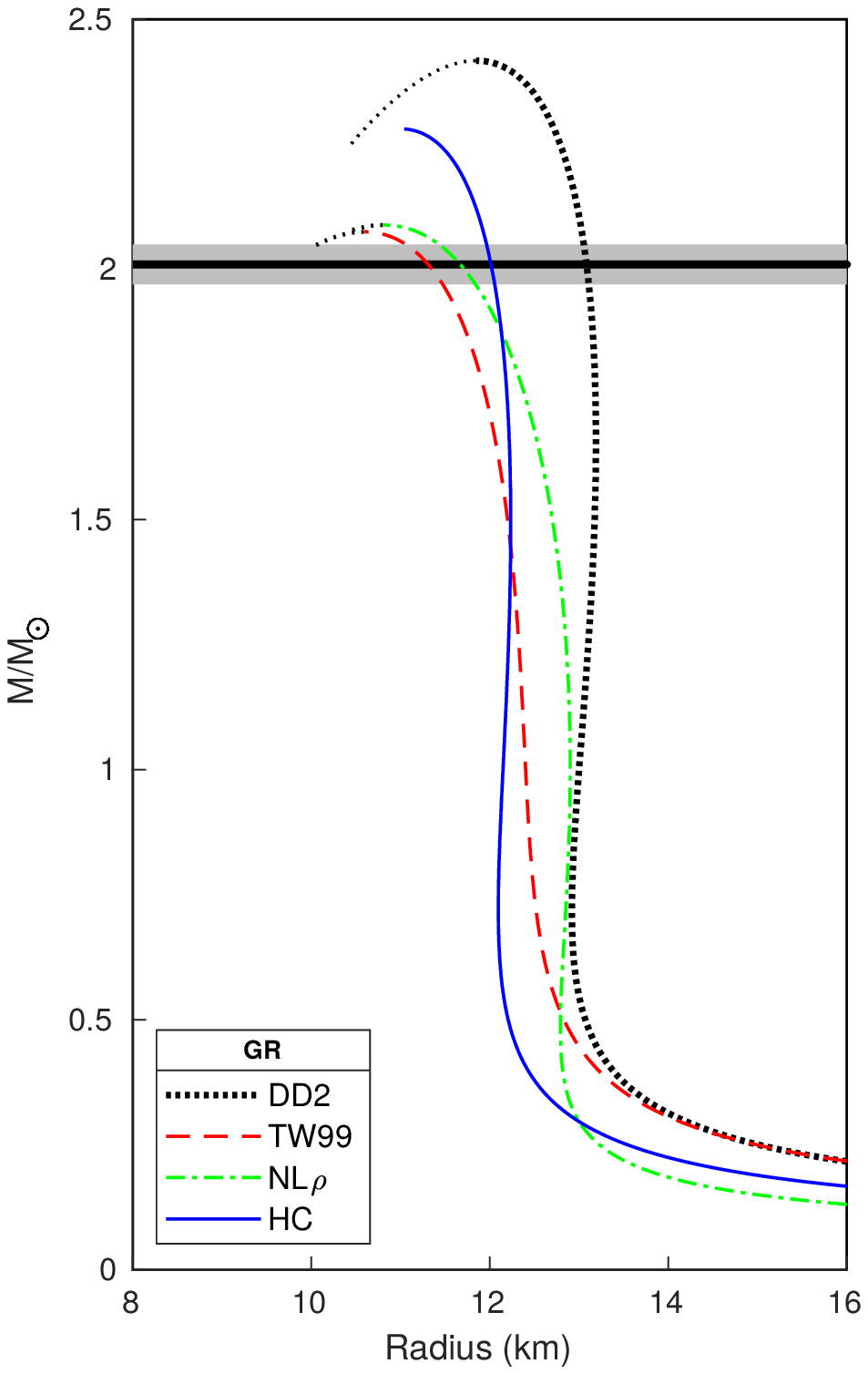}
  \mbox{(b)}
  \end{minipage}
  \caption{a)Nucleonic EoS for different models  b) Corresponding GR ($\alpha = 0$)  mass-radius sequences for the nucleonic EoSs.  The solid line corresponds to a stable configuration, beyond this the dotted lines show the solutions for unstable configurations $(dM/d\rho_c < 0)$. }
  \label{fig1}
  \end{figure*}
\subsection{Nuclear EoS}
Ideally, the dense matter in the NS core consists of neutron, protons, and electrons. We consider the EoS generated within the framework of density-dependent relativistic mean field model (DDRMF) \cite{hempel2010statistical, char2014massive}. The density-dependent couplings give rise to the rearrangement term in the pressure term, and account for the energy-momentum conservation and thermodynamic consistency of the system \cite{banik2002density}. The DDRMF model satisfies the constraints on nuclear symmetry energy and its slope parameter as well as the incompressibility from the nuclear physics experiments \cite{char2014massive}. A proper core-crust matching is very crucial to avoid uncertainties in the macroscopic properties of the stars as emphasised in Ref \cite{fortin2016neutron}. The DDRMF EoS employs the same Lagrangian density to describe low-density crust as well as the high-density core and allows a smooth transition between the core and crust. 

TW99\cite{dhiman2007nonrotating}, is another density-dependent RMF model where the Lagrangian density consists of self- and mixed-interaction terms for $\sigma$, $\omega$, and $\rho$ mesons up to the quartic order.  The self interaction of the $\sigma$ meson significantly improves the value of the nuclear matter incompressibility, whereas that of  $\omega$ mesons plays an important role in varying the high-density behavior of the EoS, and also prevents instabilities. The $\rho$-meson interaction terms accounts for the density dependence of the symmetry energy coefficient and the neutron skin thickness in heavy nuclei over a wide range without affecting other properties of finite nuclei.

From the other category with constant couplings, we consider HC, a FSU-type nucleonic RMF EoS, that uses scalar-isoscalar $\sigma$, vector-isoscalar $\omega$ meson together with their self-interactions, vector-isovector  $\rho$ meson with its cross interaction with $\omega$ meson too, and scalar-isovector $\delta$ meson as degrees of freedom\cite{bunta2003asymmetric}. The mean-field parametrizations are obtained by using realistic and  relativistic NN interactions calculated by more fundamental Dirac-Brueckner-Hartree-Fock theory (DBHF).

The final nucleonic EoS that matches the observational constraint is NL$\rho$, with $\sigma$ self-couplings and interplay between the $\rho$ and $\delta$-mesons \cite{liu2002asymmetric}.  The main effects of the $\delta$ meson field are on symmetry energy  and its slope and curvature of the nuclear system, from the EoS to n,p-mass splitting, and in particular on the nuclear response in unstable regions.

The EoSs, TW99, HC and  NL$\rho$ use the outer crust adopting the Baym, Pethick and Sutherland (BPS) model\cite{baym1971ground}.

\begin{figure*}[!t]
\begin{minipage}[b]{0.45\textwidth}
  \includegraphics[width=\linewidth, height=10cm]{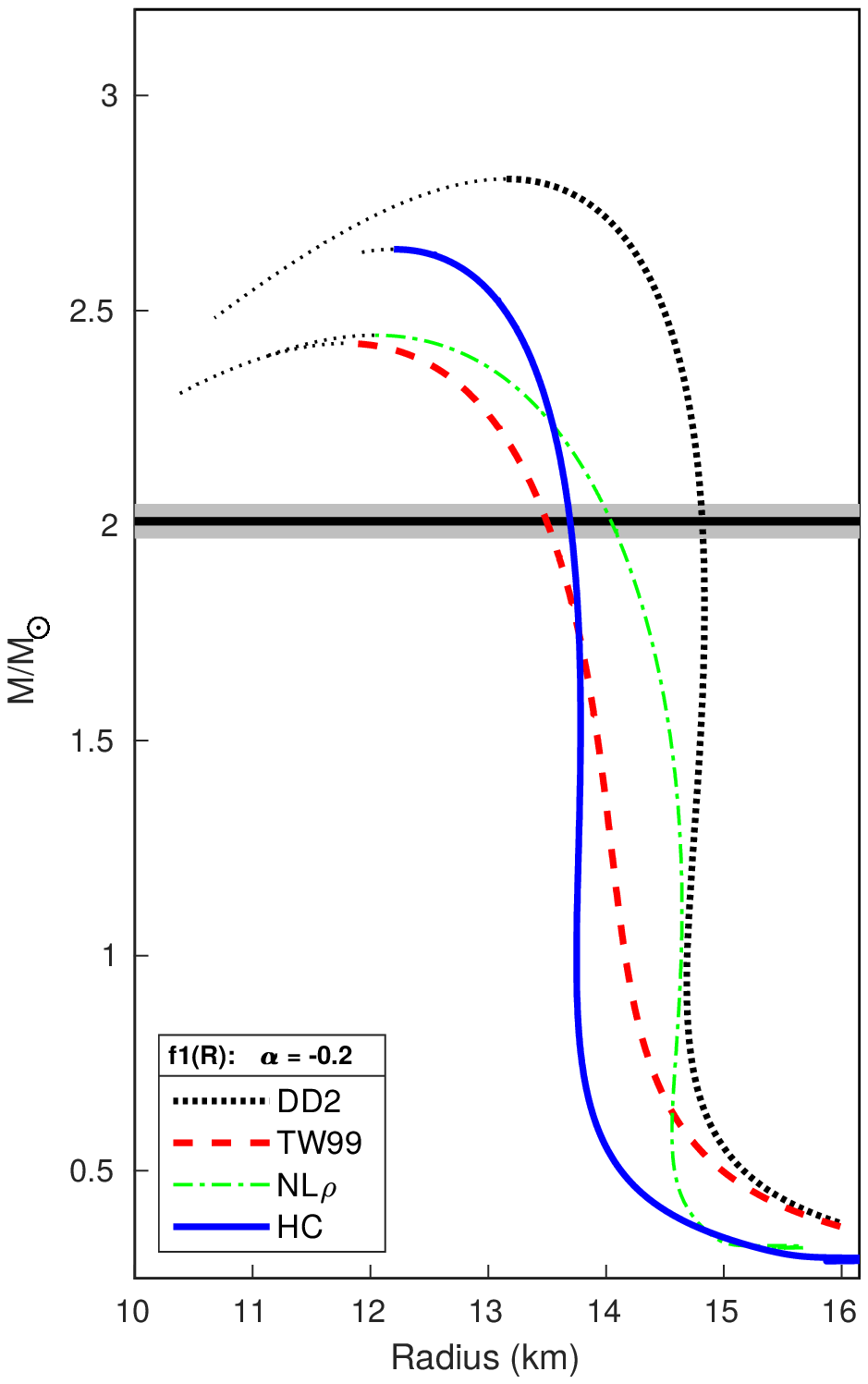}
  \mbox{(a)}
  \end{minipage}\hfill
\begin{minipage}[b]{.45\textwidth}
  \includegraphics[width=\linewidth, height=10cm]{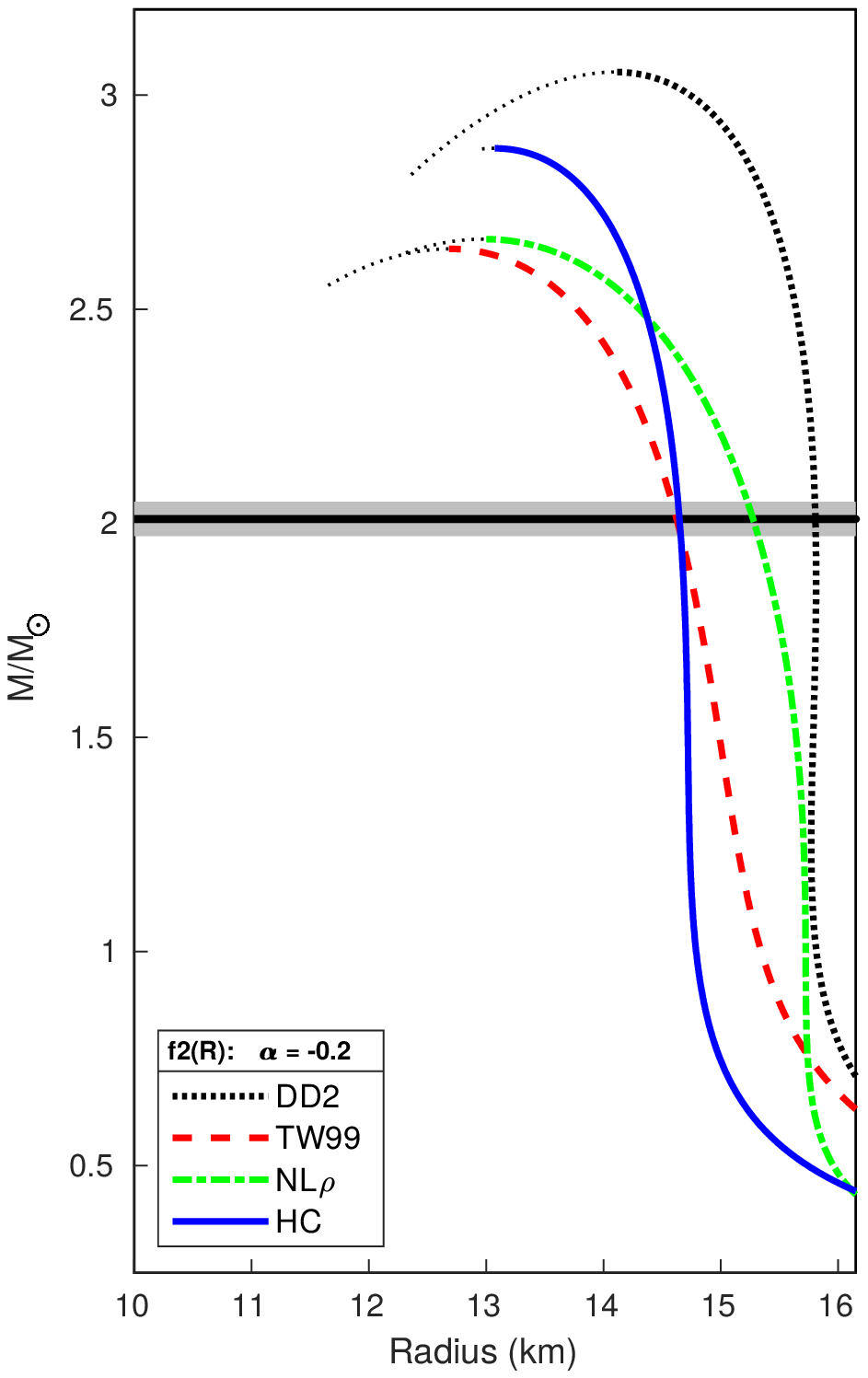}
  \mbox{(b)}
  \end{minipage}
  \caption{ The NS mass sequences with radius are plotted for nucleonic EoSs for a) $\alpha$ = -0.2 using  $f_1(R)$ model and b) $\alpha$ = -0.2 $km^2$ $f_2(R)$ model.}
  \label{fig2}
  \end{figure*}

\subsection{Strange EoS}
The high-density core of the NS is hypothesized to be populated with strange baryons and even quark matter under tremendous pressure. The strange matter might be more stable than the nucleons-only stars.  The onset of a new degree of freedom softens the EoS and supports a star that is less massive and less compact compared to the ones comprising of only nucleons. Here, we consider two particular EoSs consisting of $\Lambda$ hyperons, and quark matter in addition to the nucleons. 

We include the lightest baryon $\Lambda$ hyperons in the before-mentioned DDRMF nucleonic model, where the repulsive $\Lambda-\Lambda$ interaction is mediated via  $\phi$ mesons. The scalar meson coupling to $\Lambda$ hyperons  are determined from their potential depth (-30MeV) in normal nuclear matter, obtained from the experimental data of the single particle spectra of $\Lambda$ hypernuclei. The EoS of $\beta$-equilibrated and charge neutral cold NS matter are calculated from the $BHB\Lambda \phi$ supernova EoS Tables\cite{banik2014new}. Other baryons of the octet are not considered mainly due to the uncertainty in their experimental data.

Finally, we consider a phase transition from hadron to quark matter in the NS interior, where the quark phase is described within the phenomenological MIT bag model\cite {MIT}. The Bag constant B can be interpreted as an inward pressure needed to confine the quarks into a bag, and is a parameter that can be varied to make the EoS consistent with the stringent bound of tidal deformability and maximum mass constraint\cite{nandi2019constraining}. For the hadronic part, we choose FSWGarnet, a relativistic density functional that supports a $2 M_{\odot}$ NS\cite{fonseca2016nanograv, antoniadis2013t, demorest5788two}, and also predicts values for the symmetry energy and its slope at saturation density within experimental limits \cite{chen2015compactness}. We use Bag Constant $B^{1/4}$ =150MeV for quark matter with BPS crust \cite{baym1971ground}.

\section{Numerical Model \& Results}
Spherically symmetric configurations are obtained solving the modified TOV equations for the model EoS, namely Eqs. \ref{TOV1}, and \ref{TOV2}, for the realistic EoSs discussed above. This implies the integration from the centre to the surface of NS.  For a chosen central pressure P(r=0), with boundary conditions at the center [m(r=0)=0] and on the surface [P(r=R)=0], total mass [M=m(R)] enclosed in the star of radius R is integrated. We numerically integrate Eqs. \ref{TOV1} and \ref{TOV2} using the Euler's method with a step size of 0.001 km and employ piece-wise interpolation to obtain the intermediate energy densities. The mass and corresponding radius for each entry of the EoS are calculated to plot the corresponding M-R relations. This procedure is repeated for the range of $\alpha$s and the EoSs used in our study.

It has been shown \cite{Arapo_lu_2011} that observational constraints of 2$M_{\odot}$ from the PSR J1614-2230 excludes many of the possible soft EoS if one presumes GR as the only classical theory of gravity. In the gravity model employed, the perturbation expansion parameter $\alpha$ introduces a new degree of freedom to the theory, such that it allows some of the EoSs, which are excluded within the framework of GR, to be reconciled with the observations for definite values of $\alpha$. It is to be noted that the dimension of $\alpha$ is dependent on the model of gravity we choose. For $f_1(R)$ model $\alpha$ is dimensionless, whereas in case of $f_2(R)$, $\alpha$ has the units of $r_g^2$, and is therefore of the orders of $\approx  km^2$. We have considered both positive and negative $\alpha$ values in this analysis. For $f_2(R)$ model, $\alpha < 0$ leads to ghost and tachyonic instabilities. For a stable solution one need to match the Schwarzschild solution which is the natural exterior solution for the gravity models. For this, we must ensure that the Ricci scalar and its normal derivative vanish at the surface. These conditions make the fluid pressure vanish at the surface of the star. For the polytropic EoS one can choose the boundary conditions such that we can match with Schwarzschild even if $\alpha < 0$, which is obviously excluded because of the ghost condition\cite{Ganguly}. Our approach will also be pragmatic in a sense such that we find some stable solutions with $\alpha< 0$, as also considered earlier by some authors \cite{cooney2010neutron, Arapo_lu_2011, Cheoun, Deliduman}. In the following paragraphs, we discuss the numerical analysis for both the categories of EoS~ {\it viz:} Nuclear and Strange discussed above.
\begin{figure*}[!t]
\begin{minipage}[b]{0.45\textwidth}
  \includegraphics[width=\linewidth, height=10cm]{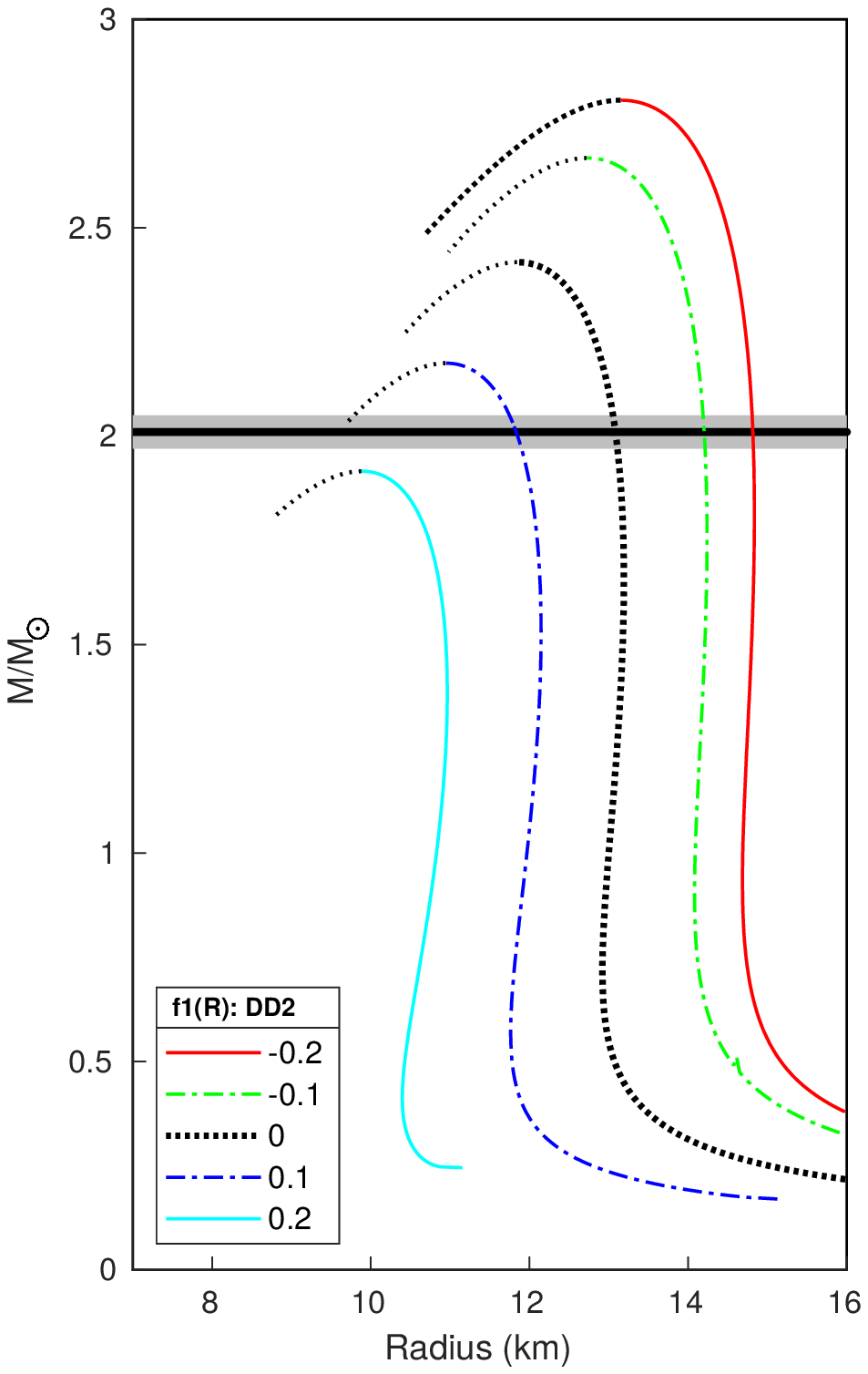}
  \mbox{(a)}
  \end{minipage}\hfill
\begin{minipage}[b]{.45\textwidth}
  \includegraphics[width=\linewidth, height=10cm]{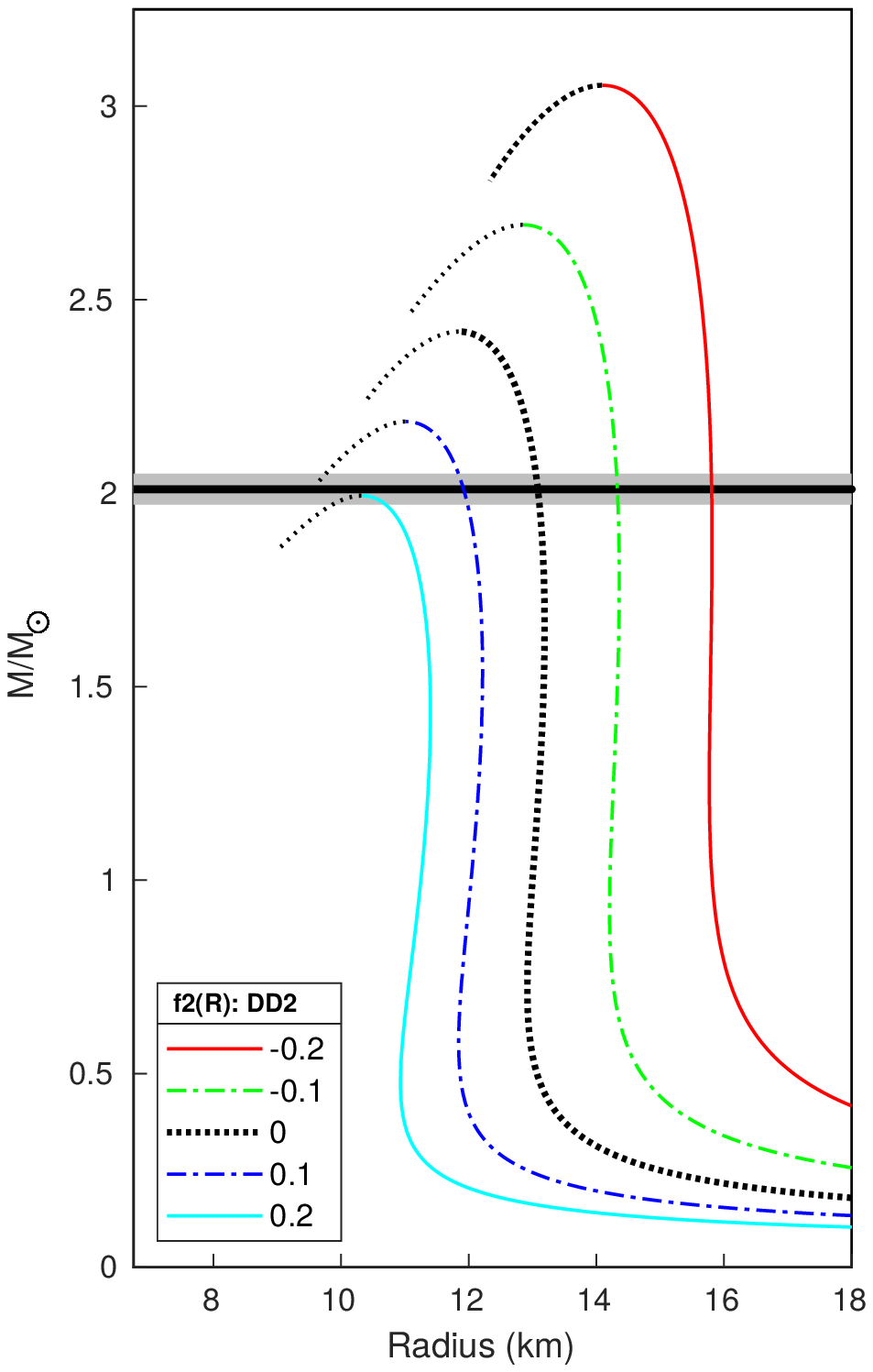}
  \mbox{(b)}
  \end{minipage}
  \caption{ The neutron star mass sequences with radius are plotted for nuclear EoS = DD2 for different values of $\alpha$ in a) $f_1(R)$ model and  b) $f_2(R)$ model.}
  \label{fig3}
\end{figure*}

\begin{table*}[t]
\caption{\label{tab:tab1}%
Maximum mass and corresponding radius of the NS for $\alpha$=0 (GR); $\alpha$ = -0.2 for $f_1(R)$ and $f_2(R)$
}
\begin{tabular}{|l|cc|cc|cc|}
\hline
Model&\multicolumn{2}{c|}{GR} &
\multicolumn{2}{c|}{$f_1(R)$} &
\multicolumn{2}{c|}{$f_2(R)$} \\
    &$M_{max}$&Radius &$M_{max}$& Radius&$M_{max}$& Radius\\
 \vspace{1mm}
    EoS &$M_{\odot}$&km &$M_{\odot}$&km    &$M_{\odot}$&km\\
    \hline
    DD2&2.417&11.846&2.806&13.156&3.054&14.094\\
    TW99&2.075&10.594&2.424&11.777&2.641&12.684\\
    NL$\rho$&2.089&10.807&2.443&12.039&2.664&12.983\\
    HC&2.281&11.043&2.642&12.200&2.876&13.073\\
    FSU&2.067&11.665&2.447&12.973&2.679&13.931\\
    \hline
    BHB$\Lambda \phi$&2.094&11.506&2.469&12.851&2.700&13.852\\
    FSUQ&2.001&11.445&2.374&12.725&2.599&13.650\\
\hline
\end{tabular}
\end{table*}

\subsection {Nuclear Eos:}

The nucleonic EoSs considered in this work, i.e., DD2, TW99, NL$\rho$ and HC, are plotted in Fig.~\ref{fig1}(a) and their mass and radius profiles are plotted in Fig.~\ref{fig1}(b) for the case of GR. From the $M-R$ plot, we notice that the maximum mass in these models are consistent with the observation of $2 M_{\odot}$  \cite{antoniadis2013t, demorest5788two}. Beyond the maximum mass, the configurations become unstable where $\frac {dM}{dr}>0.$ In these figures, the maximum mass corresponds to the point where the thick line ends.  The observational constraint on mass is shown as the horizontal black line with its error shown in grey.\\

In $f(R)$ gravity model, any viable combination of EoS and $\alpha$ must yield a $M-R$ relation with a maximum mass exceeding this measured mass. We compare the mass profiles generated in GR and the two gravity models, $f_1(R)$ and $f_2(R)$ in Fig. \ref{fig2}(a) and Fig. \ref{fig2}(b) respectively. For this, we consider our chosen set of nucleonic EoSs and calculate the maximum mass and its corresponding radius for a particular value of $\alpha=-0.2$.  Compared to GR (Fig.~\ref{fig1}(b)), the stars can support more mass in the gravity models than the standard GR for each of the EoS.\\ 


The maximum mass for each EoS model in both $f_1(R)$ and $f_2(R)$, that can be stabilised by the degeneracy pressure of the constituent fermions against gravitational collapse, are  listed in Table~\ref{tab:tab1} along with the corresponding radius. All of them are well above $2M_{\odot}$, consistent with the maximum mass of NS observed \cite{antoniadis2013t, demorest5788two}. It is observed that between the two gravity models, the maximum mass and radius of $f_1(R)$ model are closer to GR compared to that of $f_2(R)$ model.\\

Next, we study the effect of the parameter $\alpha$ on the M-R relations. In Figs.~\ref{fig3}, we show the effect of $\alpha$ on the mass-radius profile for the two f(R) gravity models using the DD2 EoS. The maximum masses and the corresponding radii for the set of $\alpha =-0.2$ to $0.2$  are tabulated in Table~\ref{tab:tab3} and \ref{tab:tab4}. The pattern observed in Figs.~\ref{fig2} for $\alpha=-0.2$ changes for other values of $\alpha$s. We notice that $\alpha>0.1$  and $\alpha>0.2$  $km^2$ is not permissible for $f_1(R)$ and  $f_2(R)$ respectively from observational constraints. A model with positive $\alpha$ values can not accommodate enough mass as supported by the GR models.\\

\begin{figure*}[!t]
\begin{minipage}[b]{0.45\textwidth}
  \includegraphics[width=\linewidth, height=10cm]{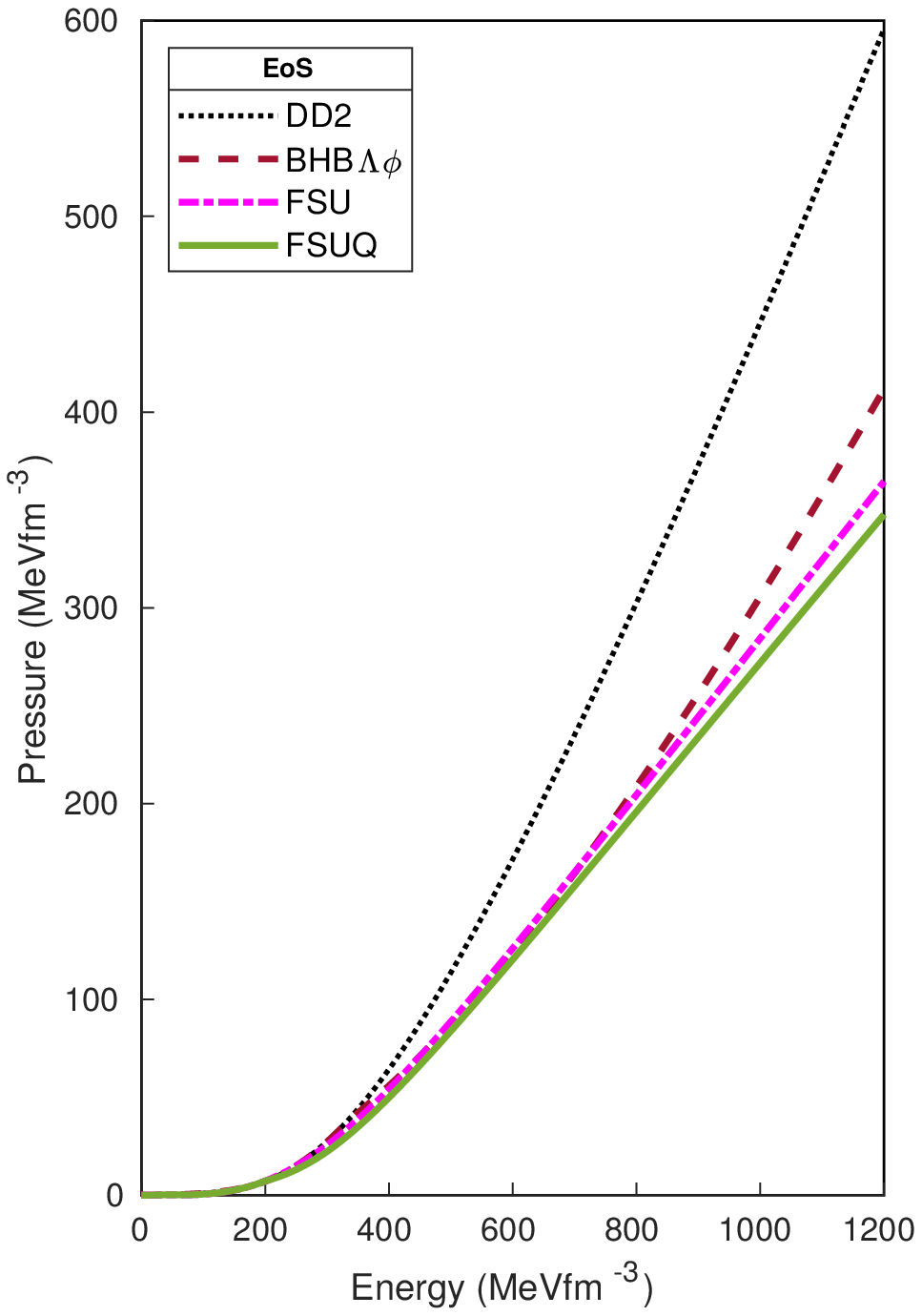}
  \mbox{(a)}
  \end{minipage}\quad
\begin{minipage}[b]{.45\textwidth}
  \includegraphics[width=\linewidth, height=10cm]{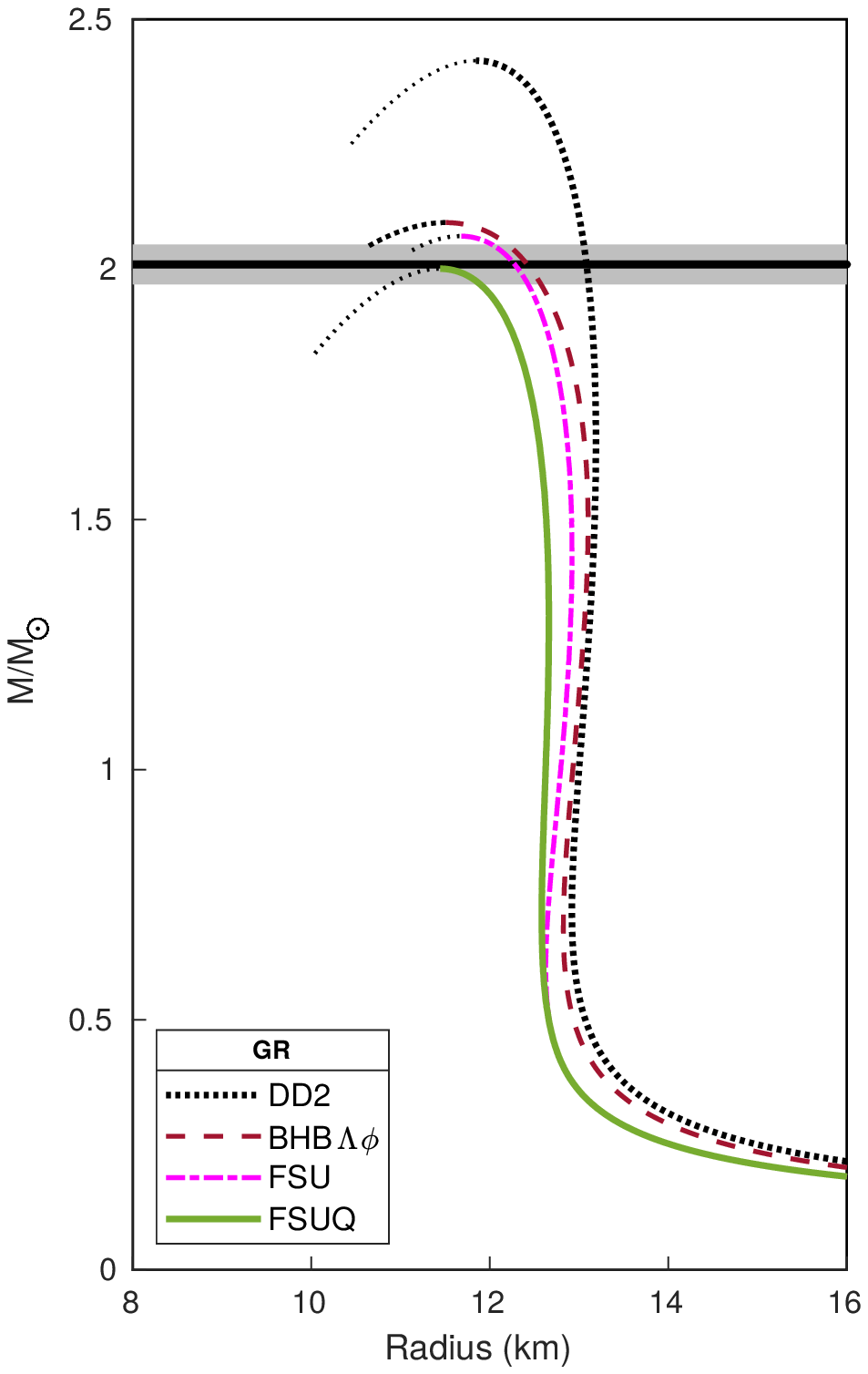}
  \mbox{(b)}
  \end{minipage}
  \caption{a) Strange EoS for different models  b) Corresponding GR ($\alpha$ = 0) mass-radius sequences for the  strange EoSs along  with their nuclear version.}
  \label{fig4}
  \end{figure*}

A similar pattern is also noted for the radius of NS. An $f(R)$ model with negative $\alpha$ leads to a bigger star compared to those with positive $\alpha$s.  
Next, we try to constrain the value of $\alpha$ based on radius measurements. The calculations on tidal deformability has set a range of radius for the NS. For a NS of $1.4M_{\odot}$, the radius is constrained to within $12 - 13.45$ km\cite{most2018new}. The NICER result for PSR J0030+0451 pulsar sets a limit for radius ($11.52 - 15.11$ km) and mass ($1.18 - 1.62M_{\odot}$) from two different groups using different models \cite{Miller_2019, 2019ApJ...887L..22R}.  These results support the $\alpha$ values in the range of -0.2 to 0.1 for DD2 with $f_{1}$(R) and -0.1 to 0.2 $km^2$ for DD2 with $f_{2}$(R) (See Table~\ref{tab:tab2}). For rest of the nuclear EoS considered, the results from these constraints are also shown in Table-\ref{tab:tab2}. Notice that for nuclear EoSs FSU, NL$\rho$  and TW99, positive values of $\alpha$ are not within the observational limit.

\begin{table}
\caption{\label{tab:tab2}%
Maximum mass and corresponding radius of the NS for different $\alpha$'s in $f_1(R)$ and $f_2(R)$ satisfying the observational constraint
}
\begin{tabular}{|l|cc|cc|}
\hline
&\multicolumn{2}{c|}{$f_1(R)$} &
\multicolumn{2}{c|}{$f_2(R)$} \\
&$M_{max}$&Radius&$M_{max}$&Radius\\
&$M_{\odot}$&km &$M_{\odot}$&km\\
\hline
    EoS &Min $\alpha$ &Max $\alpha$ &Min $\alpha (km^2)$&Max $\alpha (km^2)$\\
    \hline
    BHB$\Lambda \phi$&-0.2&0&-0.1&0\\
        FSUQ&-0.2&0&-0.1&0\\
\hline
    FSU&-0.2&0&-0.1&0\\
      NL$\rho$&-0.2&0&-0.1&0\\
       TW99&-0.2&0&-0.2&0\\
       DD2&-0.2&0.1&-0.1&0.2\\
        HC&-0.2&0.1&-0.2&0.1\\
\hline
   \end{tabular}
\end{table}

\newpage

\begin{figure*}[!t]
\begin{minipage}[b]{0.45\textwidth}
  \includegraphics[width=\linewidth, height=10cm]{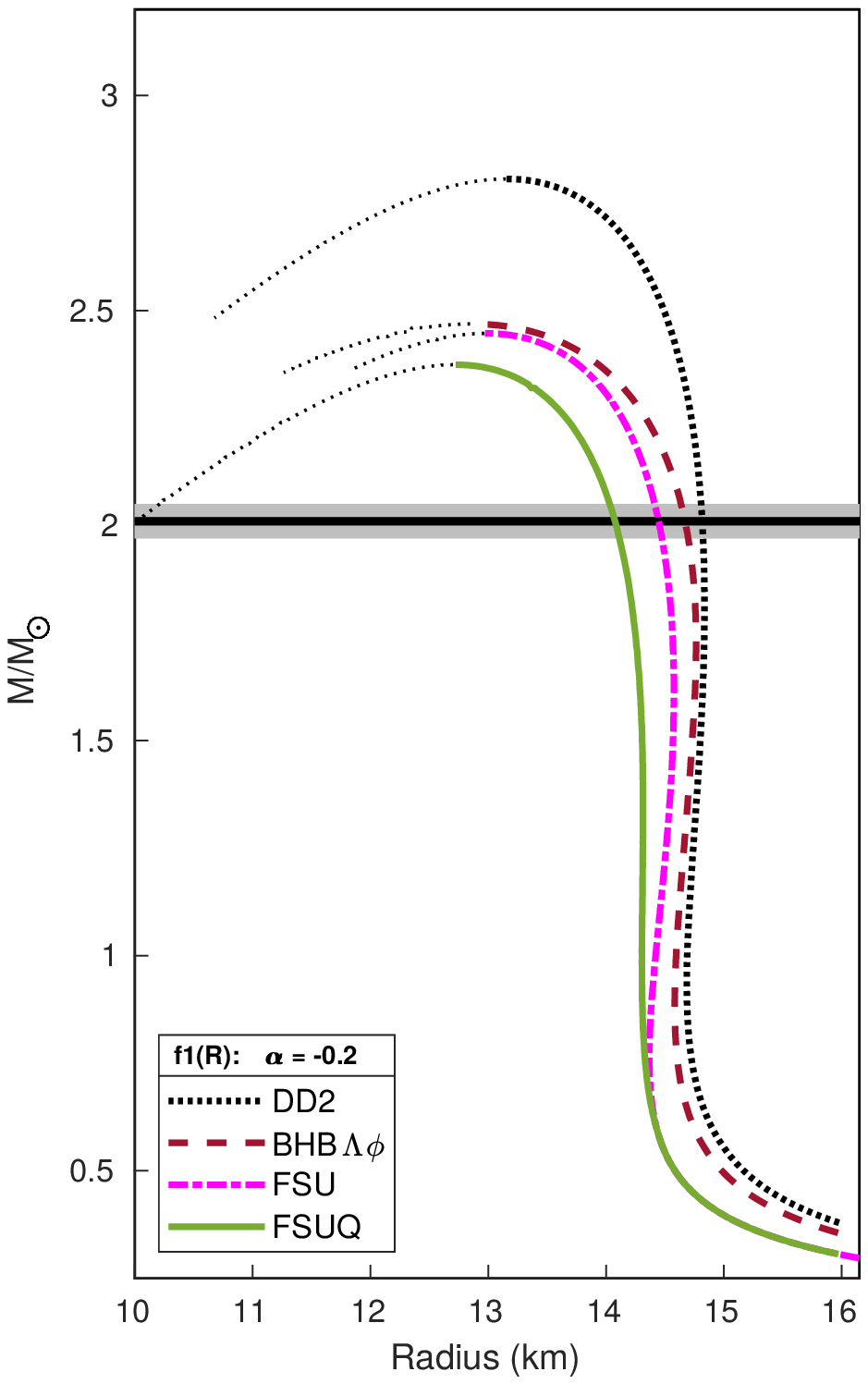}
  \mbox{\bf (a)}
  \end{minipage}\hfill
\begin{minipage}[b]{.45\textwidth}
  \includegraphics[width=\linewidth, height=10cm]{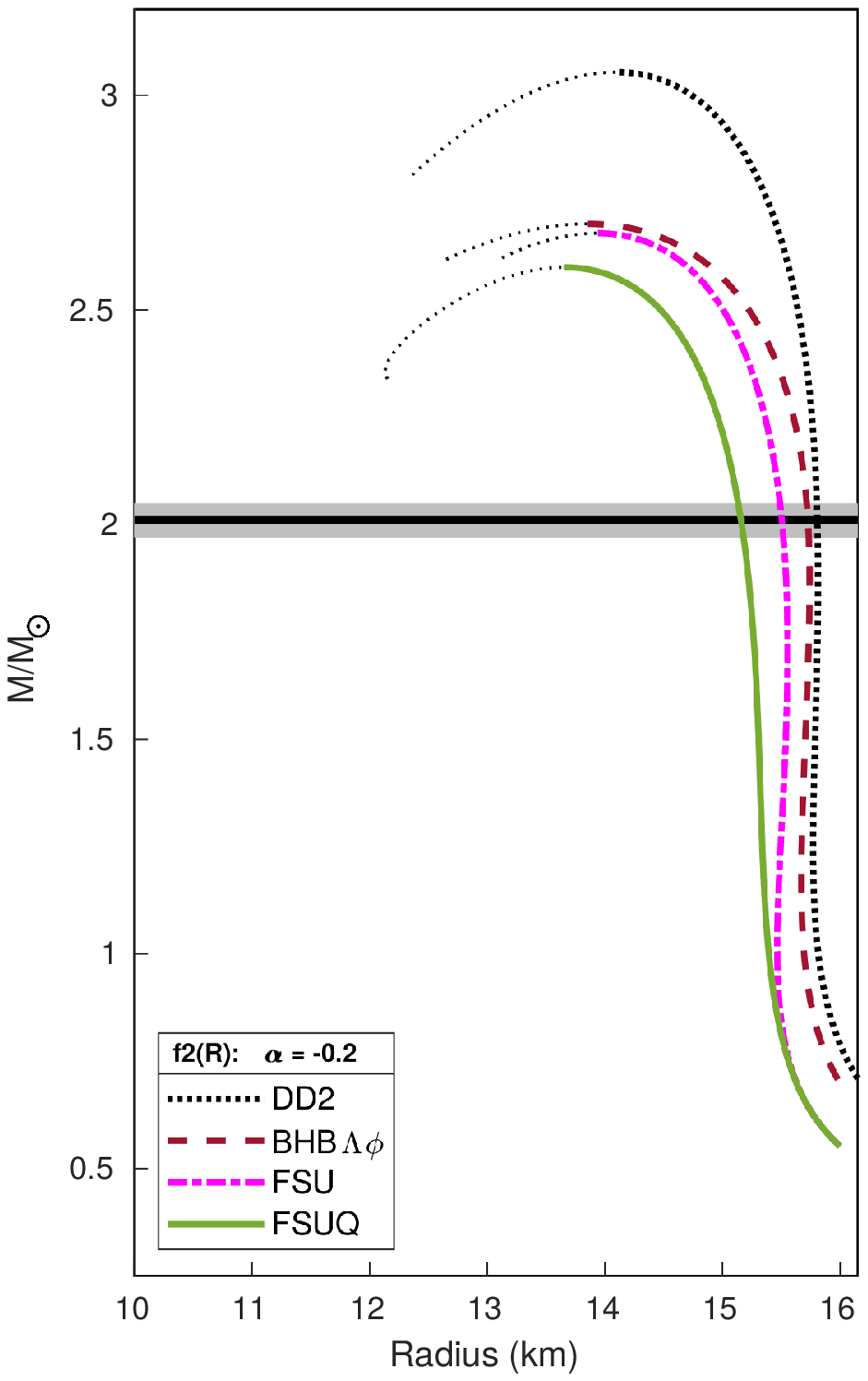}
  \mbox{\bf (b)}
  \end{minipage}
  \caption{Mass-radius sequences for strange EoSs along their nuclear version in a) $f_1(R)$ with $\alpha$ = -0.2 and  b) $f_2(R)$ with $\alpha$ = -0.2 $km^2$.}
  \label{fig5}
  \begin{minipage}[b]{0.45\textwidth}
  \includegraphics[width=\linewidth, height=10cm]{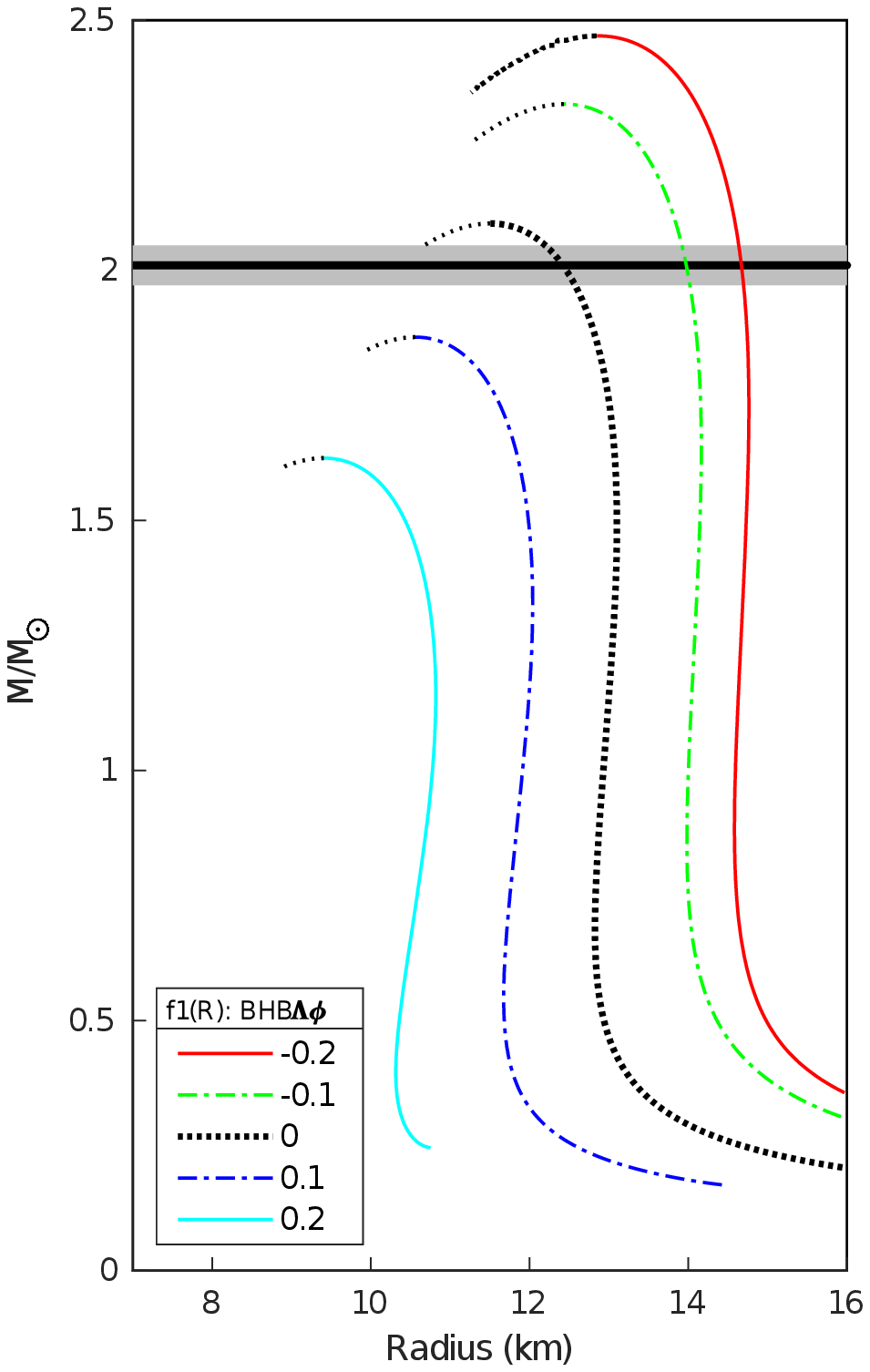}
  \mbox{\bf (a)}
  \end{minipage}\hfill
\begin{minipage}[b]{.45\textwidth}
  \includegraphics[width=\linewidth, height=10cm]{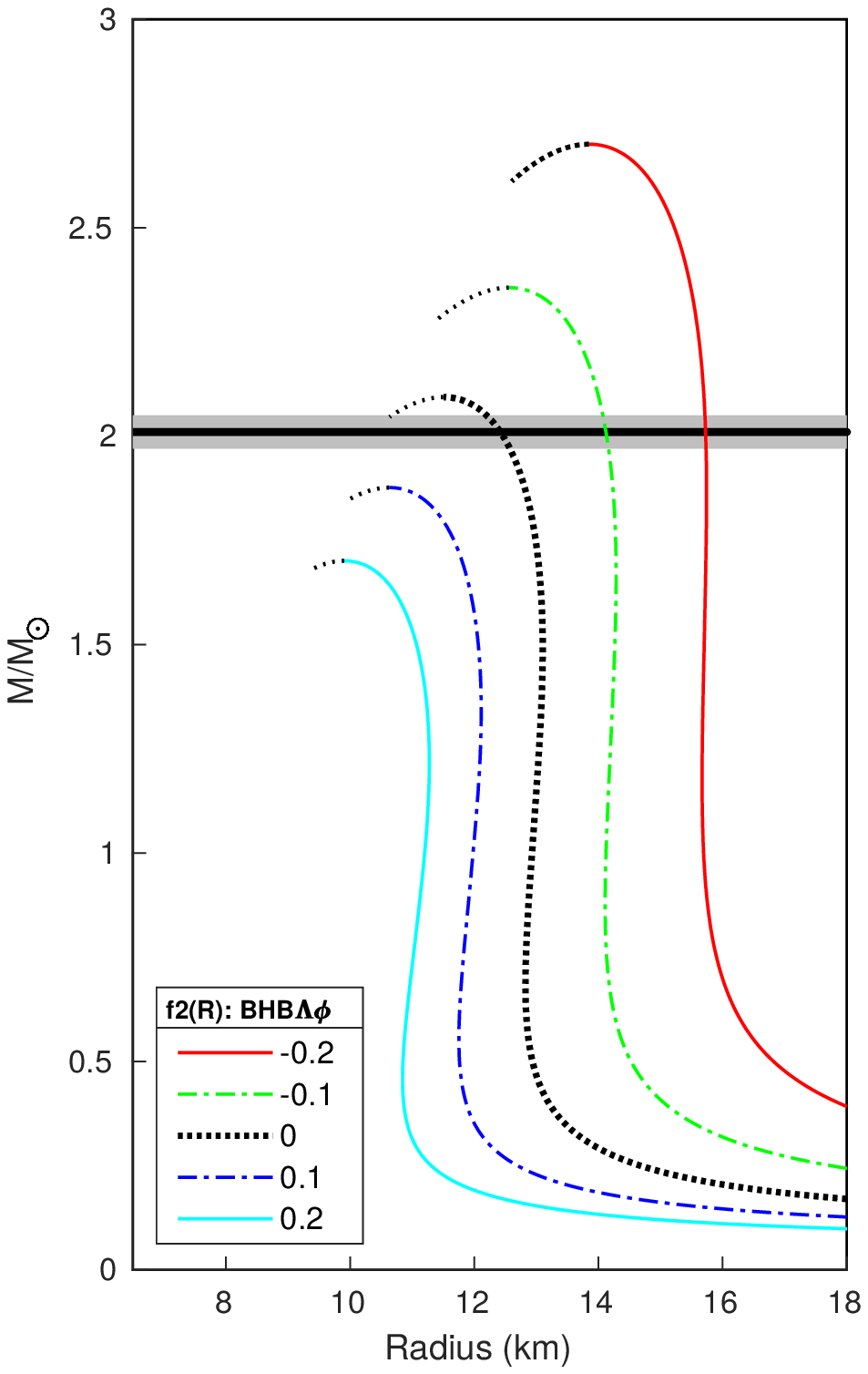}
   \mbox{\bf (b)}
  \end{minipage}
  \caption{The neutron star mass sequences with radius are plotted for EoS = BHB$\Lambda\phi$ for diffrent values of $\alpha$  using\\ a) $f_1(R)$ model. b) $f_2(R)$ model.}
  \label{fig6}
\end{figure*}

\begin{figure*}[!t]
\centering
\begin{minipage}[b]{0.45\textwidth}
\centering
  \includegraphics[width=\linewidth, height=10cm]{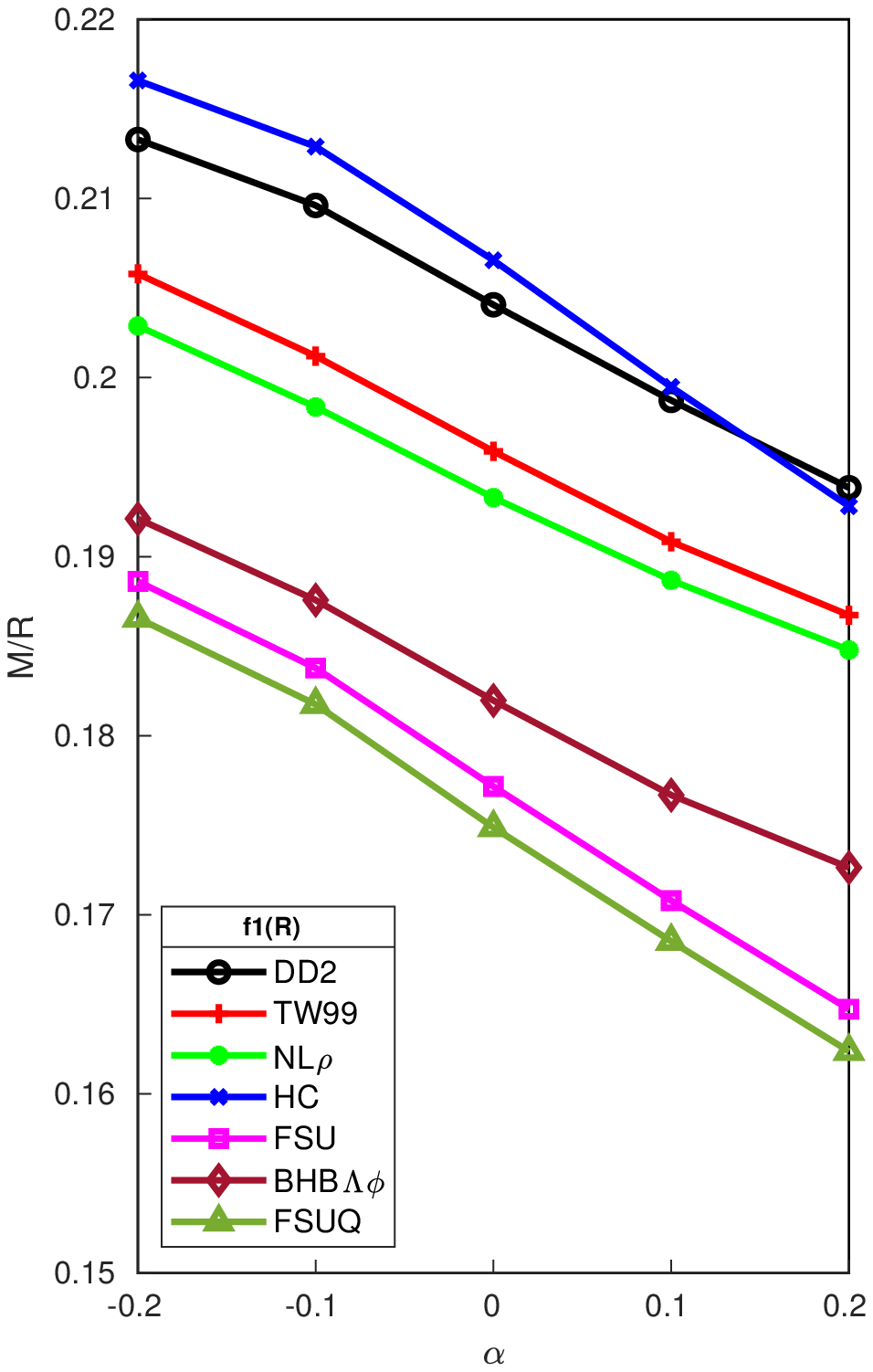}
  \mbox{\bf (a)}
  \end{minipage}\hfill
  \centering
\begin{minipage}[b]{.45\textwidth}
\centering
  \includegraphics[width=\linewidth, height=10cm]{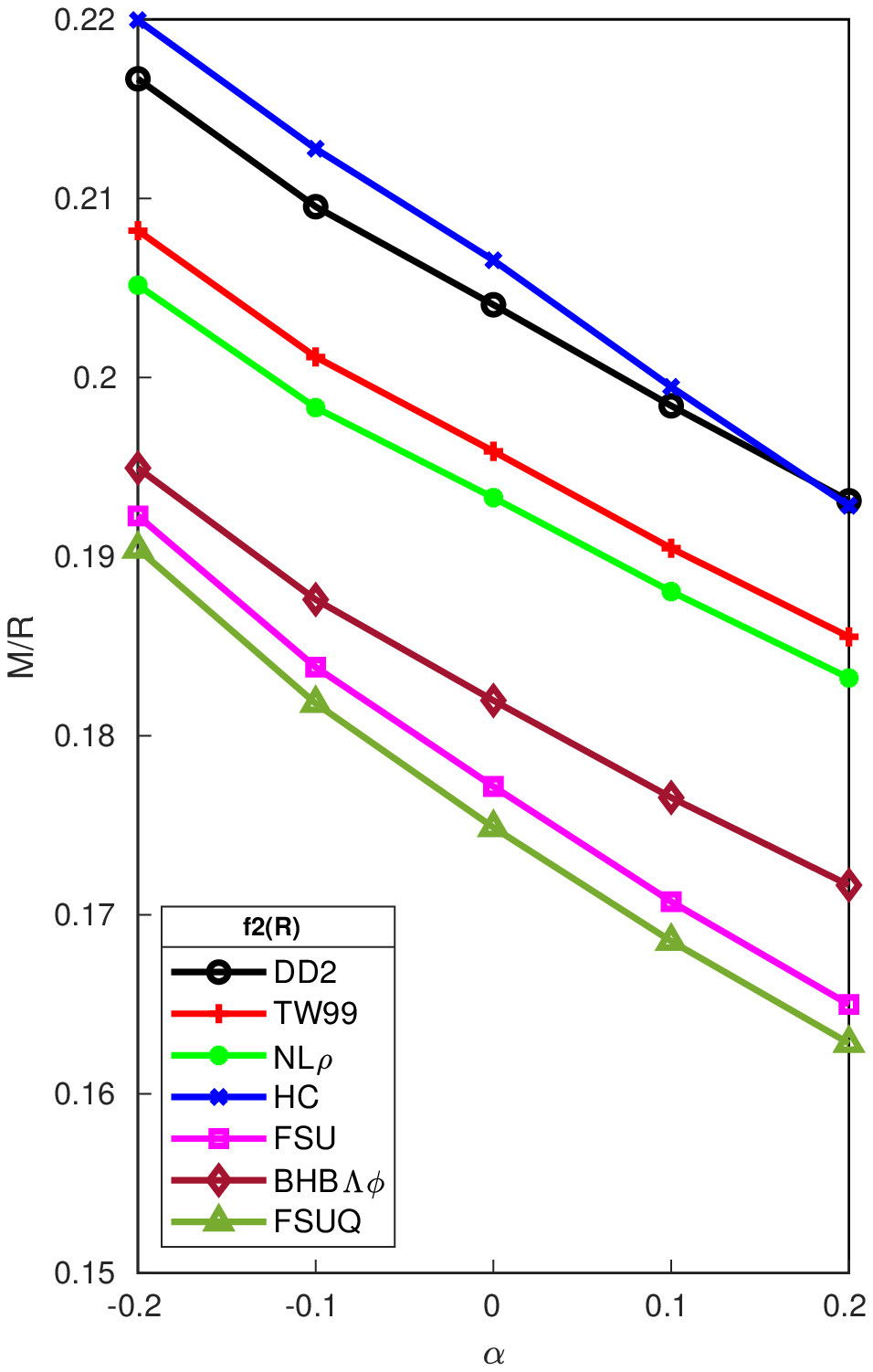}
  \mbox{\bf (b)}
  \end{minipage}
  \caption{a) M/R vs $\alpha$ for all EoSs in $f_1(R)$.  b) M/R vs $\alpha$ for all EoSs in $f_2(R)$}
  
  \label{fig7}
  \bigskip
  \centering
\begin{minipage}[b]{0.45\textwidth}
\centering
  \includegraphics[width=\linewidth, height=10cm]{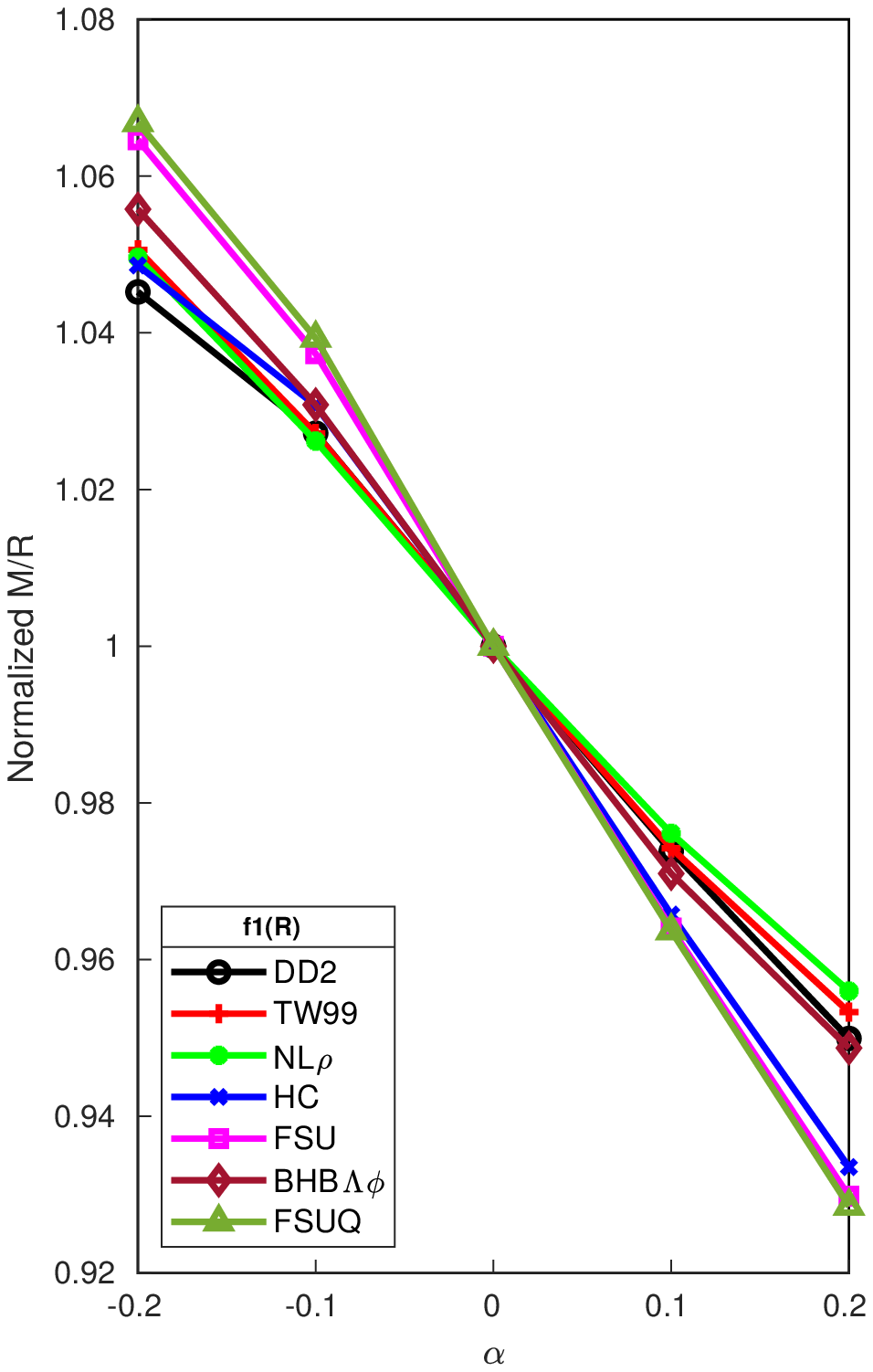}
  \mbox{\bf (a)}
  \end{minipage}\hfill
\begin{minipage}[b]{.45\textwidth}
\centering
 \includegraphics[width=\linewidth, height=10cm]{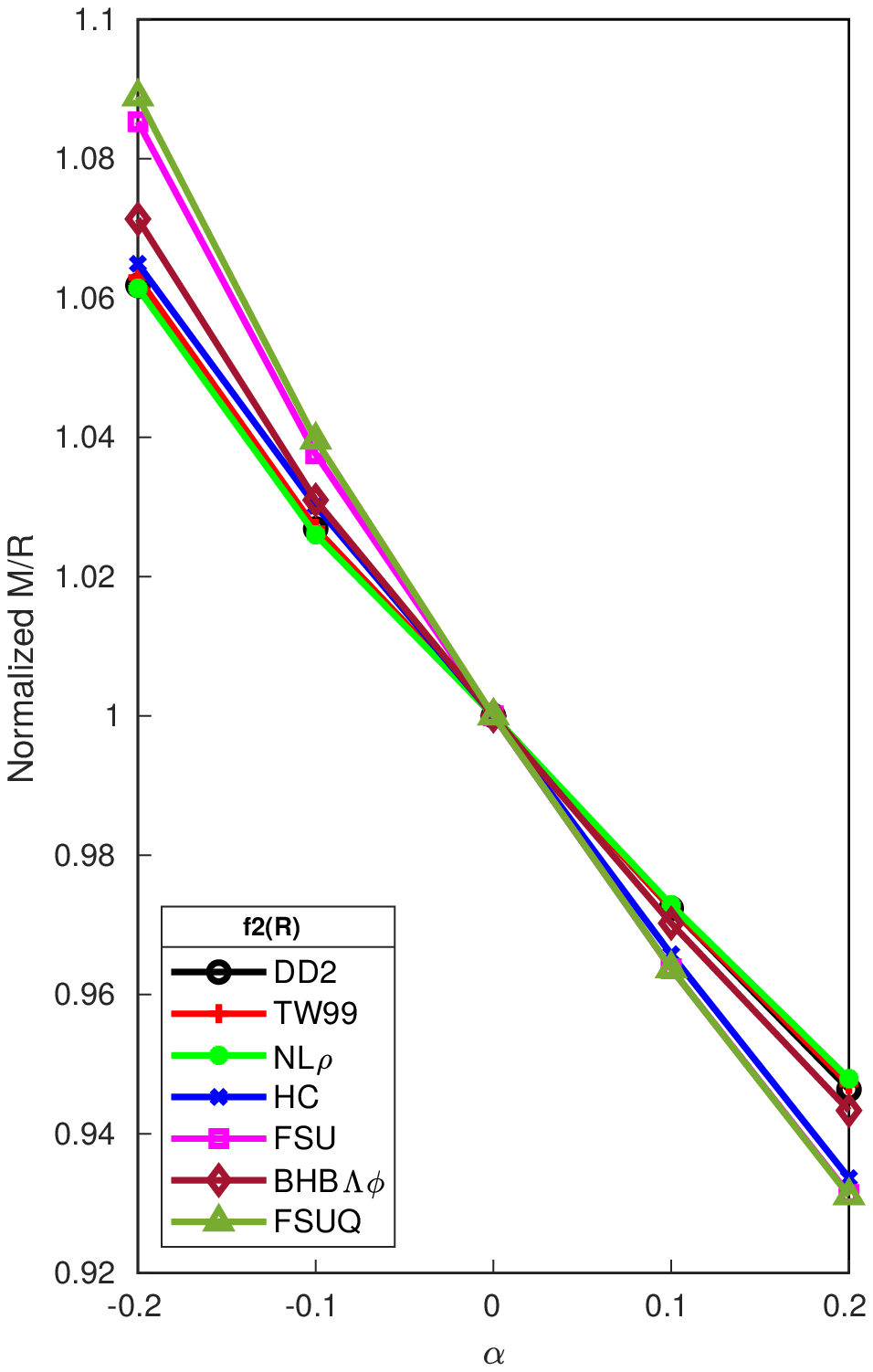}
  \mbox{\bf (b)}
  \end{minipage}
  \caption{a) Normalized M/R vs $\alpha$ for all EoSs in $f_1(R)$.  b) Normalized M/R vs $\alpha$ for all EoSs in $f_2(R)$}
  \label{fig8}
  \vfill
\end{figure*}

\begin{table*}
\caption{\label{tab:tab3}Maximum mass and corresponding radius of the NS for different values of $\alpha$'s in $f_1(R)$ gravity model.}
\begin{tabular}{|l|cc|cc|cc|cc|cc|}
\hline
     \multicolumn{1}{|c|}{$\alpha$} &
     \multicolumn{2}{c|}{-0.2} &
     \multicolumn{2}{c|}{-0.1} &
     \multicolumn{2}{c|}{0} &
     \multicolumn{2}{c|}{0.1} &
     \multicolumn{2}{c|}{0.2} \\
\hline
     &$M_{max}$&Radius&$M_{max}$&Radius&$M_{max}$&Radius&$M_{max}$&Radius&$M_{max}$&Radius\\
     &$M_{\odot}$&km&$M_{\odot}$& km&$M_{\odot}$& km&$M_{\odot}$&km&$M_{\odot}$&km\\
\hline
DD2&2.806&13.156&2.667&12.725&2.417&11.846&2.175&10.945&1.916&9.883\\
TW99&2.424&11.777&2.296&11.411&2.075&10.594&1.863&9.763&1.639&8.776\\
NL$\rho$&2.443&12.039&2.313&11.662&2.089&10.807&1.874&9.931&1.647&8.912\\
HC&2.642&12.200&2.513&11.805&2.281&11.043&2.053&10.293&1.810&9.385\\
FSU&2.447&12.973&2.309&12.563&2.067&11.665&1.836&10.749&1.587&9.631\\
BHB$\Lambda \phi$&2.469&12.851&2.332&12.435&2.094&11.506&1.866&10.563&1.625&9.410\\
FSUQ&2.374&12.725&2.238&12.314&2.001&11.445&1.776&10.541&1.533&9.444\\
\hline
\end{tabular}
\end{table*}

\begin{table*}
\caption{\label{tab:tab4}Maximum mass and corresponding radius of the NS for different values of $\alpha$ in $f_2(R)$ gravity model.}
\begin{tabular}{|l|cc|cc|cc|cc|cc|}
\hline
     \multicolumn{1}{|c|}{$\alpha (km^2)$} &
     \multicolumn{2}{c|}{-0.2} &
     \multicolumn{2}{c|}{-0.1} &
     \multicolumn{2}{c|}{0} &
     \multicolumn{2}{c|}{0.1} &
     \multicolumn{2}{c|}{0.2} \\
\hline
     &$M_{max}$&Radius&$M_{max}$&Radius&$M_{max}$&Radius&$M_{max}$&Radius&$M_{max}$&Radius\\
     &$M_{\odot}$&km&$M_{\odot}$& km&$M_{\odot}$& km&$M_{\odot}$&km&$M_{\odot}$&km\\
\hline
DD2&3.054&14.094&2.693&12.853&2.417&11.846&2.185&11.012&1.994&10.323\\
TW99&2.641&12.684&2.318&11.525&2.075&10.594&1.872&9.826&1.706&9.193\\
NL$\rho$&2.664&12.983&2.336&11.778&2.089&10.807&1.882&10.010&1.714&9.357\\
HC&2.876&13.073&2.537&11.924&2.281&11.043&2.062&10.339&1.881&9.757\\
FSU&2.679&13.931&2.333&12.690&2.067&11.665&1.847&10.817&1.668&10.112\\
BHB$\Lambda\phi$&2.700&13.852&2.356&12.560&2.094&11.506&1.877&10.630&1.701&9.909\\
FSUQ&2.599&13.650&2.261&12.438&2.001&11.445&1.787&10.605&1.613&9.909\\
\hline
\end{tabular}
\end{table*}

\begin{figure*}[!h]
\begin{minipage}[b]{0.45\textwidth}
  \includegraphics[width=\linewidth, height=10cm]{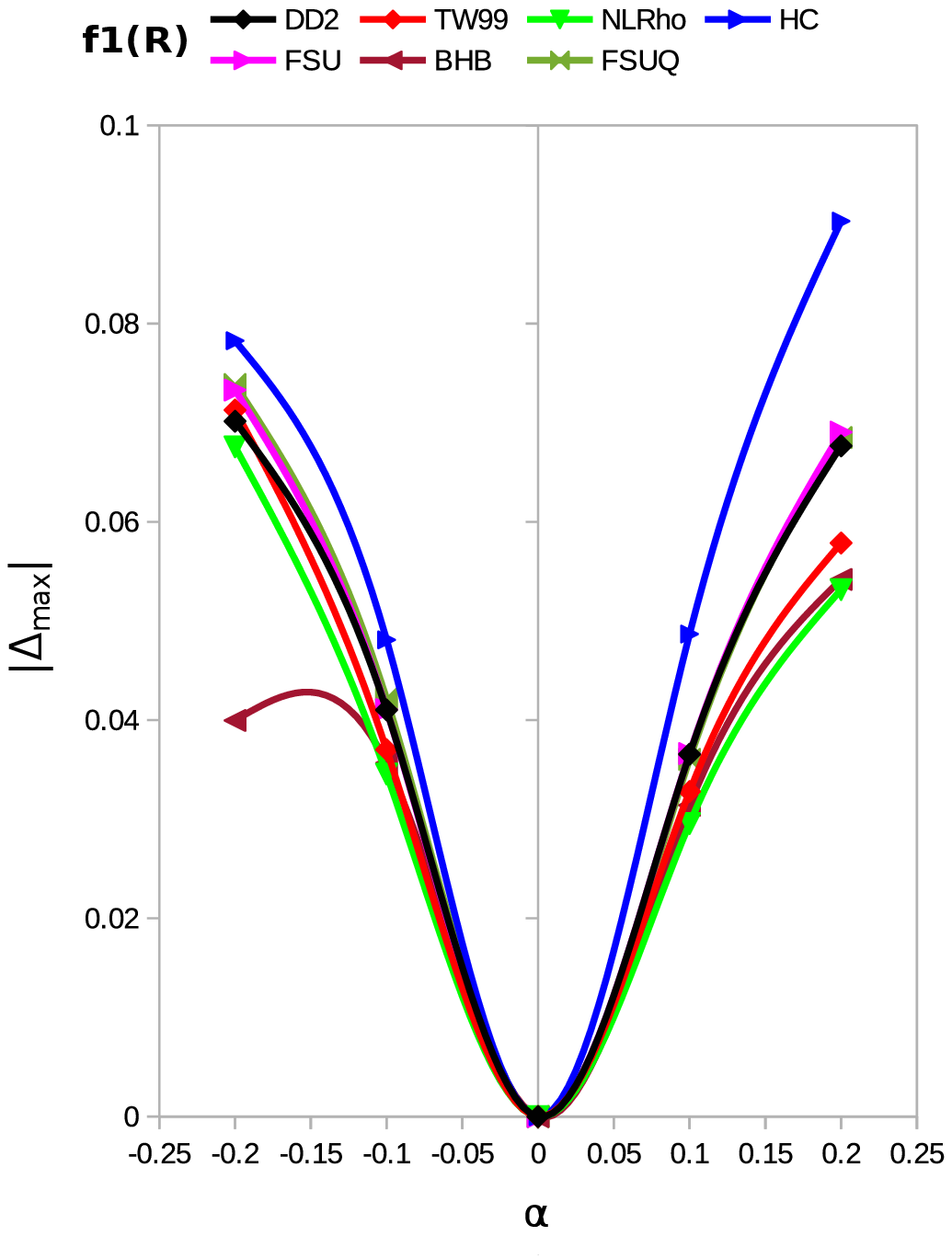}
  \end{minipage}\hfill
\begin{minipage}[b]{.5\textwidth}
  \includegraphics[width=\linewidth, height=10cm]{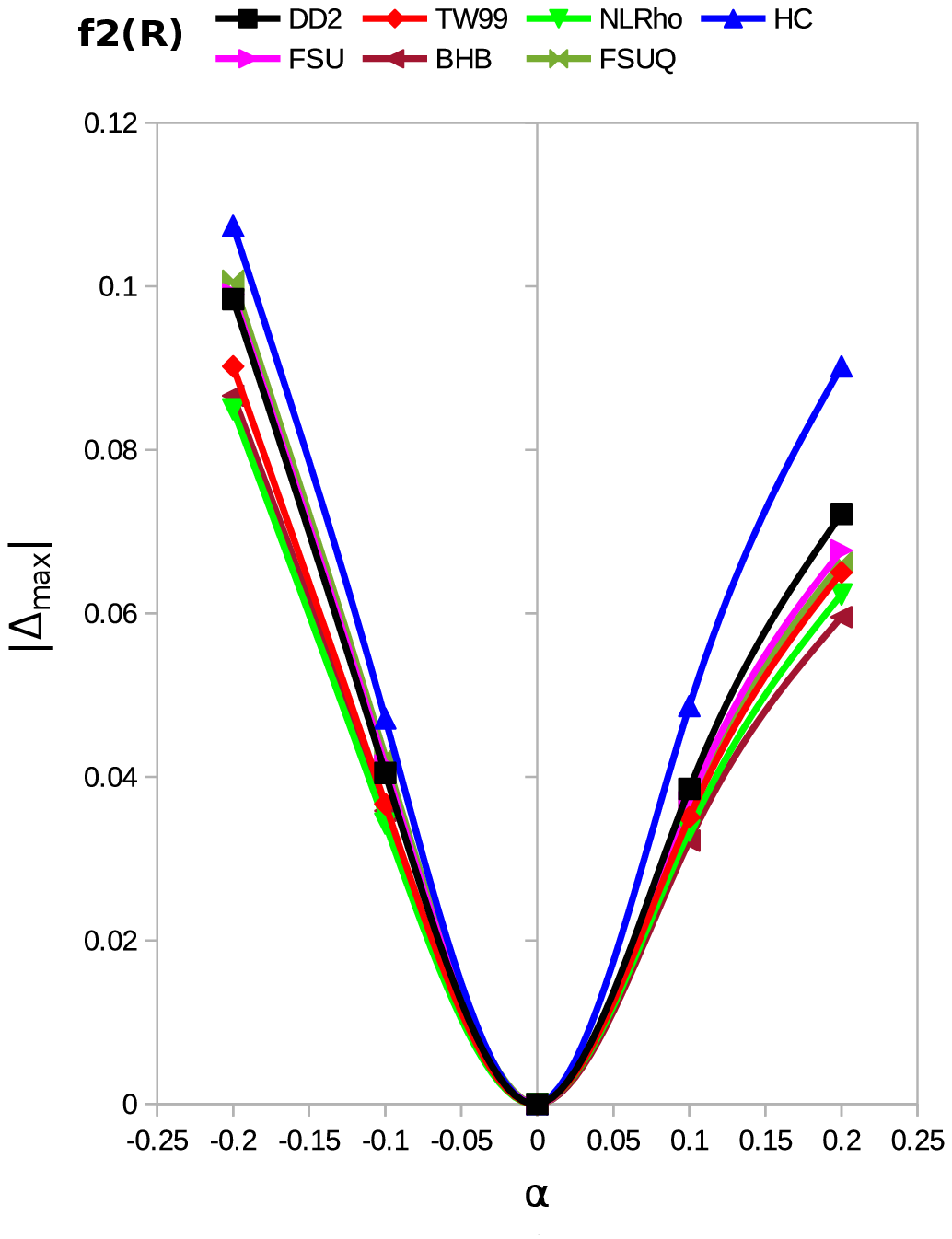}
  \end{minipage}
  \caption{ $|\Delta_{max}|$ as a function of $\alpha$ for all EoSs in a) $f_1(R)$ and  b) $f_2(R)$.}
  \label{fig9}
\end{figure*}
\subsection{Strange EoS:}
Next, we investigate the effect of strangeness on the f(R) gravity models.
For this, we consider the presence of $\Lambda$ hyperons and quarks in the dense interior of the NS. In Fig.~\ref{fig4}(a), we plot the EoSs i) BHB$\Lambda \phi$ and ii) FSUGarnet with quark matter EoS. For the sake of comparison, we also plot their nucleon-only versions. The strangeness degrees of freedom make the EoS softer. Their mass and radius profiles are plotted in Fig.~\ref{fig4}(b) for the case of GR. From the plot, it can be clearly seen that, the additional degree of freedom in strange EoS results in decrease in maximum mass than their nucleon-only counterpart. Their corresponding values are listed in Table-\ref{tab:tab1}.\\

We notice a marked difference in M-R plots of the strange EoS in case of f(R) gravity models.  Fig.~\ref{fig5}(a) and \ref{fig5}(b) depict the M-R relation of the strange EoS along with their nuclear-only version for $\alpha=-0.2$ in $f_1(R)$ and $f_2(R)$ gravity models respectively. As seen in the case of GR, the maximum mass is less but the radius is more or less same as the nuclear-only NS.
The effect of $\alpha$ on the strange BHB$\Lambda\phi$ EoS is plotted in Figs.~\ref{fig6} for both the gravity models. It shows in both the cases, the negative values of $\alpha$ yields more mass and radius as compared to GR.
\clearpage However, for positive values of $\alpha$  the maximum mass is not within the observational limit.
The permissible range of $\alpha$ from the observations of maximum mass, calculation of tidal deformity, and NICER results for each EoS considered is tabulated in Table~\ref{tab:tab2}. The maximum mass and the corresponding radius for all the EoSs discussed for a range of $\alpha$ are illustrated in Table-\ref{tab:tab3} and Table-\ref{tab:tab4} for the gravity model $f_1$(R) and $f_2$(R), respectively. In Fig. \ref{fig7}(a) and \ref{fig7}(b), we have plotted the compactness (M/R) of NS against $\alpha$ for $f_1(R)$ and $f_2(R)$ models, respectively. As the value of $\alpha$ increases from -0.2 to 0.2, the compactness decreases for all EoSs in both f(R) models, the slopes in $f_2(R)$ model being slightly greater than those in $f_1(R)$ model. Also, it is noted that stiffer the EoS, higher is the compactness of the stars. A similar behaviour can be seen in Fig. \ref{fig8}(a) and \ref{fig8}(b), where we have plotted the M/R of NS (normalized at $\alpha$=0) against $\alpha$ for $f_1(R)$ and $f_2(R)$ models, respectively.
\subsection {Perturbative regime}
 For all the cases considered here with respect to different values of $\alpha$ in the two f(R) models for the various EoS, the maximum allowed deviation from the GR prediction is constrained by the requirement that the solutions hold their perturbative validity. To be within the perturbative regime it is important that the first order corrections to the metric are small. This can be measured with

\begin{equation}
 |\Delta| = \left|\frac{A_{f(R)}(r)}{A_{GR}(r)} - 1\right|
\end{equation}
where $A_{f(R)}(r)$ and $A_{GR}(r)$ are the $rr$ component of the metric defined in Eq. \ref{mass} for f(R) and GR, respectively. This quantity is a function of radius of each star and depends on the EoS.  It has a maximum value near the core of the star, where the density and the curvature is large. As we need the entire solution to be perturbatively close to GR, we evaluate $\Delta$ at its maximum. The necessary condition for perturbative validity is that the maximum ratio $|\Delta_{max}| <1$.

Figs.~\ref{fig9}(a) and \ref{fig9}(b) show $|\Delta_{max}|$ as a function of parameter $\alpha$ for all EoSs in $f_1(R)$ and $f_2(R)$ gravity models respectively. These figures demonstrate that NS in both the gravity models can certainly be treated perturbatively as long as the magnitude of $\alpha$ is in the range $-0.2<\alpha<0.2$.

\section{Conclusion}
The study of NS is important in relativistic astrophysics for various reasons. First, they are the most stable astrophysical compact objects where matter is in extremely high-field regimes. Also, they are the key to understand the final stages of stellar evolution.  Apart from these standard roles, NS could be an important tool to test alternative theories of gravity, as huge gravitational field acts on them.
In this paper, we analyze the NS solutions with realistic EoS in perturbative $f(R)$ gravity. For this, we choose a few EoSs, generated in the framework of relativistic mean field models. The NS core contains strange particles in the form of $\Lambda$ hyperons and quarks, in addition to the nucleons and leptons. The mass-radius values for static NSs, calculated in GR by solving the TOV equations lie well within the observational limits for all the chosen EoSs. As the field equations in $f(R)$ gravity are of fourth order, obtaining modified TOV equations in a standard fashion is difficult unlike in the case of GR, which has second order field equations. 
In order to resolve this problem, we adopt a perturbative approach \cite{cooney2010neutron} for handling the corrections to GR. This formulation views the new terms as only the next to leading order terms in a larger expansion. Here, extra terms in the gravity action are multiplied by a parameter $\alpha$. The extra terms with some appropriate value of $\alpha$ act perturbatively and modify the results obtained in the case of GR. Such formalism allows us to explore alternative theories of gravity while maintaining important consistency conditions, including the conservation of stress-energy and gauge invariance. For our purpose, we use this formalism for two models of $f(R)$ gravity:$f_1(R)=R+ \alpha R(e^{-R/R_0}-1)$ and $f_2(R)=R+\alpha R^2$. 

Numerical integration of the modified TOV equations using Euler's method provides the mass-radius relation for NS with nuclear and strange EoS in f(R) gravity models. The predicted mass-radius relation for NS in both $f_1(R)$ and $f_2(R)$ gravity models differs from that computed within GR. In $f_1(R)$ gravity model, the negative values of parameter $\alpha$ results in increase of the maximum mass and radius of NS than in GR, whereas for positive $\alpha$ these values are less than GR. In the perturbative limit, the permissible range of $\alpha$ constrained from maximum mass observations, tidal deformity and NICER values for the nucleonic EoS DD2 and HC are $-0.2<\alpha<0.1$ and for FSU, NL$\rho$, TW99 and  the strange EoSs BHB$\Lambda\phi$ FSUQ are $-0.2<\alpha<0$.

In case of  $f_2(R)$, the laboratory bound from the E$\ddot{o}$t-Wash experiment provides the small bound on $\alpha (\leq 10^{-16}km^2)$, while the results from Gravity Probe B imply a much larger limit on $\alpha(\leq 5\times10^5 km^2)$. The measurements of the precision of the pulsarB in the PSR J0737-3039 system provide instead the limit on $\alpha \leq 2.3 \times 10^9 km^2$. Even for these large values of $\alpha$ the quadratic term in $f_2(R)$ still induces a small correction of GR. For the EoSs considered, we see that our results are well within these value. However, they provide more stringent constraints on the permissible range of $\alpha$. For the nucleonic EoS DD2 and HC are $-0.2 km^2<\alpha<0.2 km^2$ and for FSU, NL$\rho$, TW99 and the strange EoSs BHB$\Lambda\phi$ FSUQ are $-0.2 km^2<\alpha<0$. Further it is seen that increase in $\alpha$ in this range decreases the compactness (M/R) of NS for all EoS in both f(R) models. We show that the range of $\alpha$ we considered is within the perturbative limit.  We can thus, from our analysis, conclude  that the constraints we obtained from the realistic EoSs are actually the strongest constraints we could obtain by using NS as the experimental probe on f(R) gravity models.

\section{Acknowledgements}
Authors thank K. Y. Ek\c{s}i for useful discussions. AA is thankful to UGC for providing financial support under the
scheme Dr. D.S. Kothari postdoctoral fellowship.
\\


\end{document}